# Modelling the 3D physical structure of astrophysical sources with GASS

D. Quénard,[1,2★] S. Bottinelli[1,2] and E. Caux[1,2]

[1] Univ. Toulouse, UPS-OMP, Institut de Recherche en Astrophysique et Planétologie (IRAP), UMR F-5277 Toulouse, France
[2] CNRS, IRAP, UMR 5277, 9 Av. Colonel Roche, BP 44346, F-31028 Toulouse Cedex 4, France

## ABSTRACT

The era of interferometric observations leads to the need of a more and more precise description of physical structures and dynamics of star-forming regions, from pre-stellar cores to protoplanetary discs. The molecular emission can be traced in multiple physical components such as infalling envelopes, outflows and protoplanetary discs. To compare with the observations, a precise and complex radiative transfer modelling of these regions is needed. We present GASS (Generator of Astrophysical Sources Structure), a code that allows us to generate the three-dimensional (3D) physical structure model of astrophysical sources. From the GASS graphical interface, the user easily creates different components such as spherical envelopes, outflows and discs. The physical properties of these components are modelled thanks to dedicated graphical interfaces that display various figures in order to help the user and facilitate the modelling task. For each component, the code randomly generates points in a 3D grid with a sample probability weighted by the molecular density. The created models can be used as the physical structure input for 3D radiative transfer codes to predict the molecular line or continuum emission. An analysis of the output hyper-spectral cube given by such radiative transfer code can be made directly in GASS using the various post-treatment options implemented, such as calculation of moments or convolution with a beam. This makes GASS well suited to model and analyse both interferometric and single-dish data. This paper is focused on the results given by the association of GASS and LIME, a 3D radiative transfer code, and we show that the complex geometry observed in star-forming regions can be adequately handled by GASS+LIME.

## 1 INTRODUCTION

The ability to predict line emission is crucial in order to make a comparison with observations. Different modelling approximations and hypothesis can be considered depending on the complexity of the problem. From Local Thermodynamic Equilibrium (LTE) to full radiative transfer codes, the goal is always to predict the physical properties of the source the most accurately possible. Non-LTE calculations can be very time consuming but are often needed in most of the cases since many studied regions are far from LTE. A few freely usable codes are available such as RATRAN[1] (Hogerheijde & van der Tak 2000), a 1D radiative transfer code (a 2D version also exists upon request to the authors); LIME (Brinch & Hogerheijde 2010), a 3D one; and MC3D[2] and RADMC-3D,[3] both also 3D codes.

Among the choice of 3D radiative transfer codes available to date, LIME is the only one doing a full non-LTE ALI (*accelerated lambda iteration*) continuum and gas line radiative transfer treatment. Other available codes only offer an LTE or Large Velocity Gradient (LVG) gas line radiative transfer (RADMC-3D) or a dust continuum radiative transfer (MC3D and RADMC-3D). LIME is based on RATRAN and a benchmarking of the two codes has been made by Brinch & Hogerheijde (2010) using the method described in van Zadelhoff et al. (2002). LIME is well suited for the treatment of most physical problems due to its performance and its flexible use: proper treatment of line blending, multiple species input, multiline raytracing and multicore parallelization.

We have developed a user-friendly interface, GASS (Generator of Astrophysical Sources Structure), in order to easily define the physical structure of a star-forming region and create input models for LIME. Thanks to its interface, GASS allows us to create, manipulate and mix one or several different physical components such as spherical sources (see Section 3.1), discs (see Section 3.2) and outflows (see Section 3.3). The functioning of GASS follows three distinct parts: (1) the grid generation, where GASS generates its 'working zone' (region where the points of the grid are generated); (2) the creation of the physical models from the grid and (3) the post-treatment analysis options, created to deal with output hyper-spectral data cubes (created by LIME for instance). GASS is fully coupled to LIME but it can be easily adapted to any existing (or future) radiative transfer code. GASS is freely available for the community (upon request at the moment, website in construction[4]) as a standalone application for Mac OS X, Windows and any Unix-based operating system. A scripted MatLab version is also available.

★E-mail: quenard.david@gmail.com
[1] http://www.sron.rug.nl/~vdtak/ratran/
[2] http://www.astrophysik.uni-kiel.de/~star/index.php?seite=mc3d
[3] http://www.ita.uni-heidelberg.de/~dullemond/software/radmc-3d/

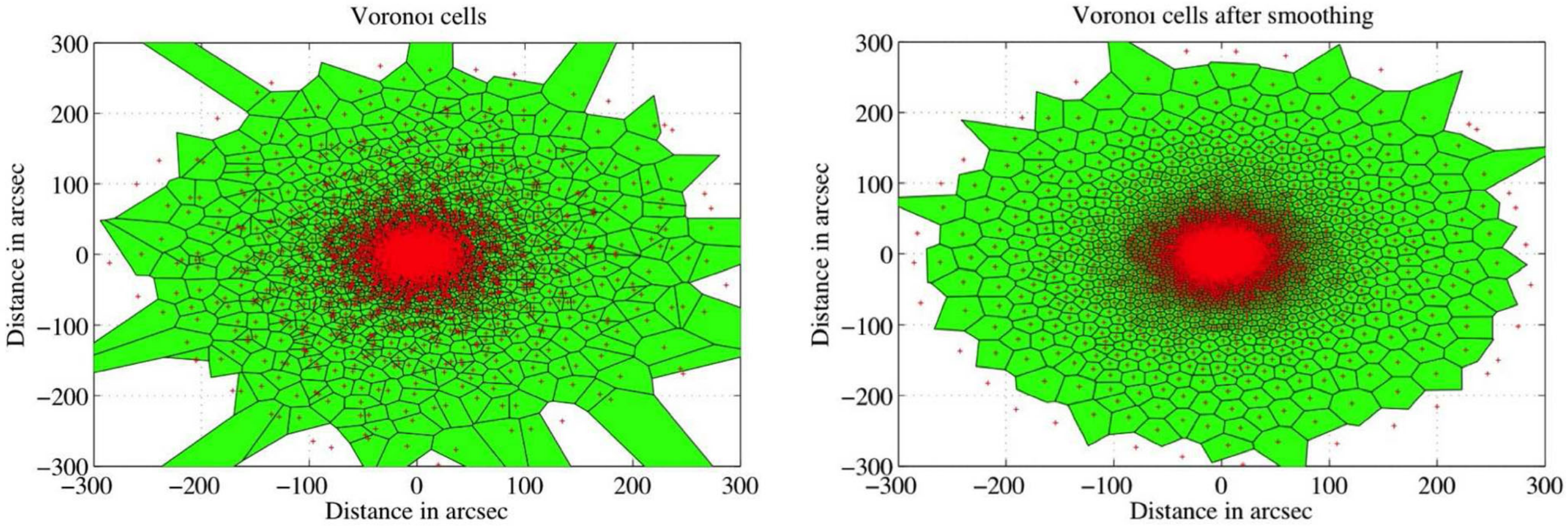

**Figure 1.** The Voronoï cells before (left) and after (right) the smoothing process with the Lloyd algorithm. The grid points are shown in red.

The outline of this paper is the following. Section 2 presents the gridding process used to create the model with GASS, Section 3 describes how the physical properties of each of the structure is generated by the code, Section 4 shows how GASS is linked to LIME, Section 5 presents the different data cube analysis options available in GASS, Section 6 gives an example of the 3D modelling capabilities of GASS, and finally, conclusions are given in Section 7.

## 2 GRID GENERATION

LIME requires models to be set in a Cartesian grid that can be defined by two means: (a) from a script called `model.c` included directly in the LIME code that allows us to define the physical properties very basically, for instance as a function of the radius without any complex 3D structure; (b) from an input model parameter file of the source in which the physical parameters (temperature, density, abundance,...) are described at each point of the Cartesian grid; this is called the pre-generated grid (or pregrid) mode. It is important to keep in mind that an input file is a huge gain in time since the model is created before giving it to LIME. Otherwise, LIME will generate its own grid as a function of the desired number of points and this step can be very time consuming.

The model generation is entirely managed by GASS and the procedure is the following.

(i) Creation of the random grid as a function of the number of points. In order to have a good convergence of the calculations, the advice is to use at least a few thousands points for the grid (Brinch & Hogerheijde 2010). Each point is randomly distributed on the grid depending on the desired structure. Points are always generated within the desired inner and outer radii of the grid. In the spherical source case, the distribution of grid points follows the density profile. Such a distribution leads to an increasing number of grid points per unit volume towards the centre to follow the distribution of the volume density across the spherical source. In the disc and outflow cases, considering their specific geometries, we decided to distribute the points equally all over their structure.

(ii) From this point distribution, the Voronoï diagram is built (see Appendix A for more explanations). Due to the random nature of the grid, two given points of the grid can be created much closer or further apart than desired. This leads to a very irregular Voronoï diagram and an insufficient number of points in the grid will produce non-homogenous effects induced by the different sizes and shapes of the Voronoï cells. A few thousands points is the minimum required to avoid this effect.

(iii) The final step consists in smoothing the grid to reduce even more the previous effect. LIME does include a smoothing process. However, in the pre-grid mode, this process is not applied (LIME only builds the Delaunay grid and the Voronoï diagram from the input grid), so it is compulsory to do it before giving the grid to LIME. We based the smoothing on the Lloyd algorithm (Lloyd 1982; Springel 2010), which consists in moving every point in the centre of mass of its Voronoï cell. From the new points positions, the Voronoï diagram is re-created and the process is repeated (items ii and iii). We have taken into account the suggestions given in section 3.1 of Brinch & Hogerheijde (2010) to determine the number of iterations and the displacement of points: indeed badly chosen values could lead to a perfectly regular grid, which would then not represent any more the underlying physical structure. In LIME, each grid point is moved slightly away from its nearest neighbour (10 per cent by iterations using 25 iterations), while in GASS each grid point is directly moved to the centre of mass of the Voronoï cell using 10 iterations. These two approaches are equivalent and they both produce a sufficiently smooth grid that preserves the underlying physical structure well.

This procedure is only adapted to describe a grid for LIME but it can easily be changed to fit the requirements of any other 3D radiative transfer code in future releases of GASS.

Fig. 1 displays the Voronoï diagram before and after the smoothing of a 2D grid. The points of the grid is plotted in red. The comparison between the two figures shows the impact of the smoothing algorithm after 10 iterations: the sizes of the Voronoï cells are more homogeneous at a given radius but the grid points are still distributed randomly.

To illustrate the different point distribution and smoothing effect according to the different structures, we performed one simulation for each of the three structures available in GASS (envelope, disc, outflow); the results are shown in Fig. 2 where the cumulated number of points in the grid as a function of the radius in arcsec are plotted. The blue and red curves are the distribution of points before and after smoothing, respectively. In the top panel, a spherical source located at 120 pc generated with 10 000 points distributed over a 50 arcsec grid was considered. In this plot, one can notice that the smoothing process only moves the points in the 3D grid without affecting the distribution as a function of the radius. During the smoothing process, some of the points will be moved inside and outside of the inner and outer edges of the computational domain (in this example: 0.1 arcsec for the inner edge and 50 arcsec for the outer one), and these points are not included in the grid anymore. To be sure that the minimum number of points will be at least around 10 000 points after rejection, 2.5 per cent more points are arbitrarily added (total of 10 250 points) during the creation of the grid. In this example, 239 points have been rejected after the smoothing process.

[4] http://quenarddavid.wixsite.com/astrophysics/gass-code

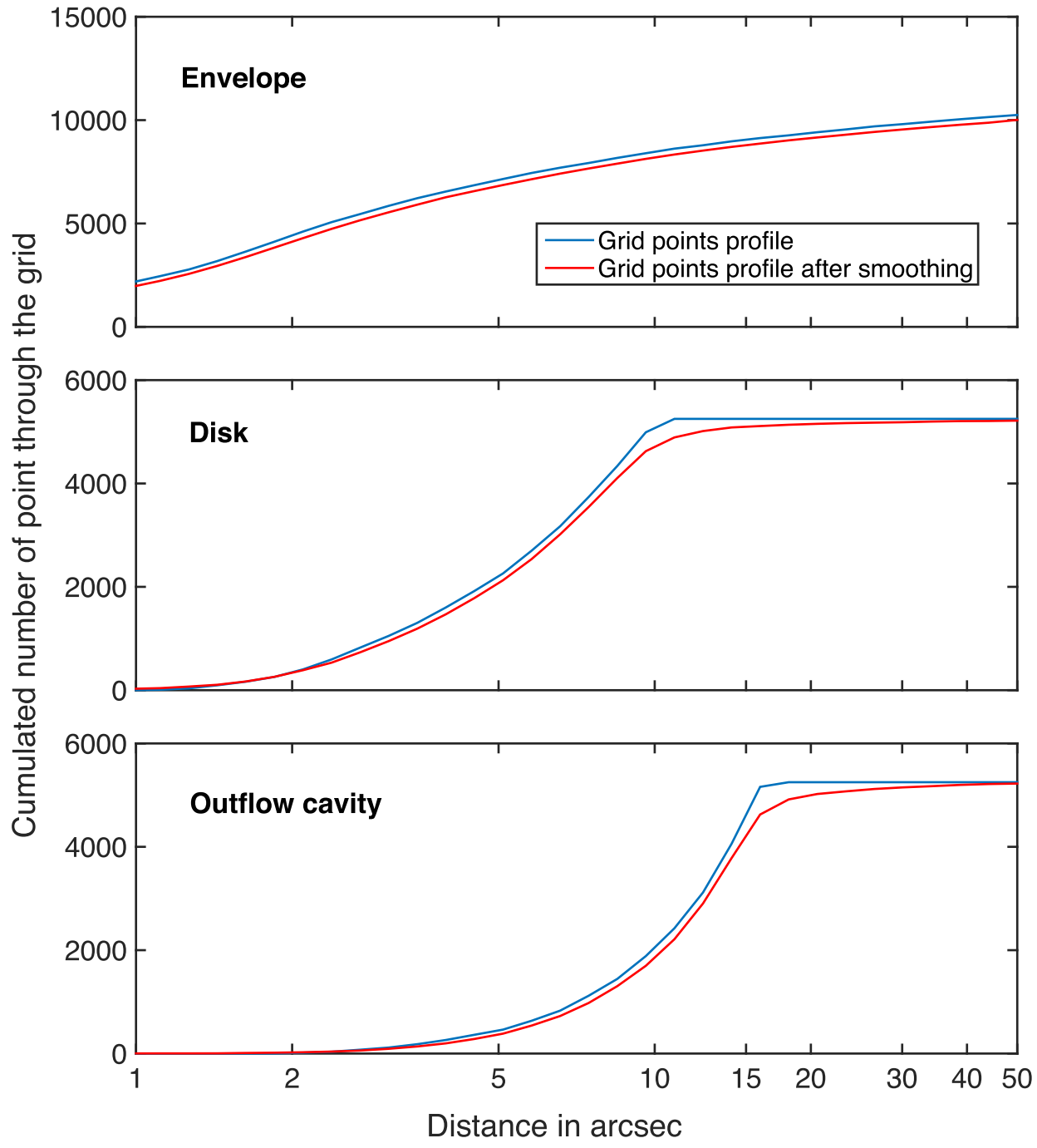

**Figure 2.** Cumulated number of points through the grid as a function of the radial distance in arcsec for the three different types of structure. The distribution of points is shown before (in blue) and after (in red) the smoothing process.

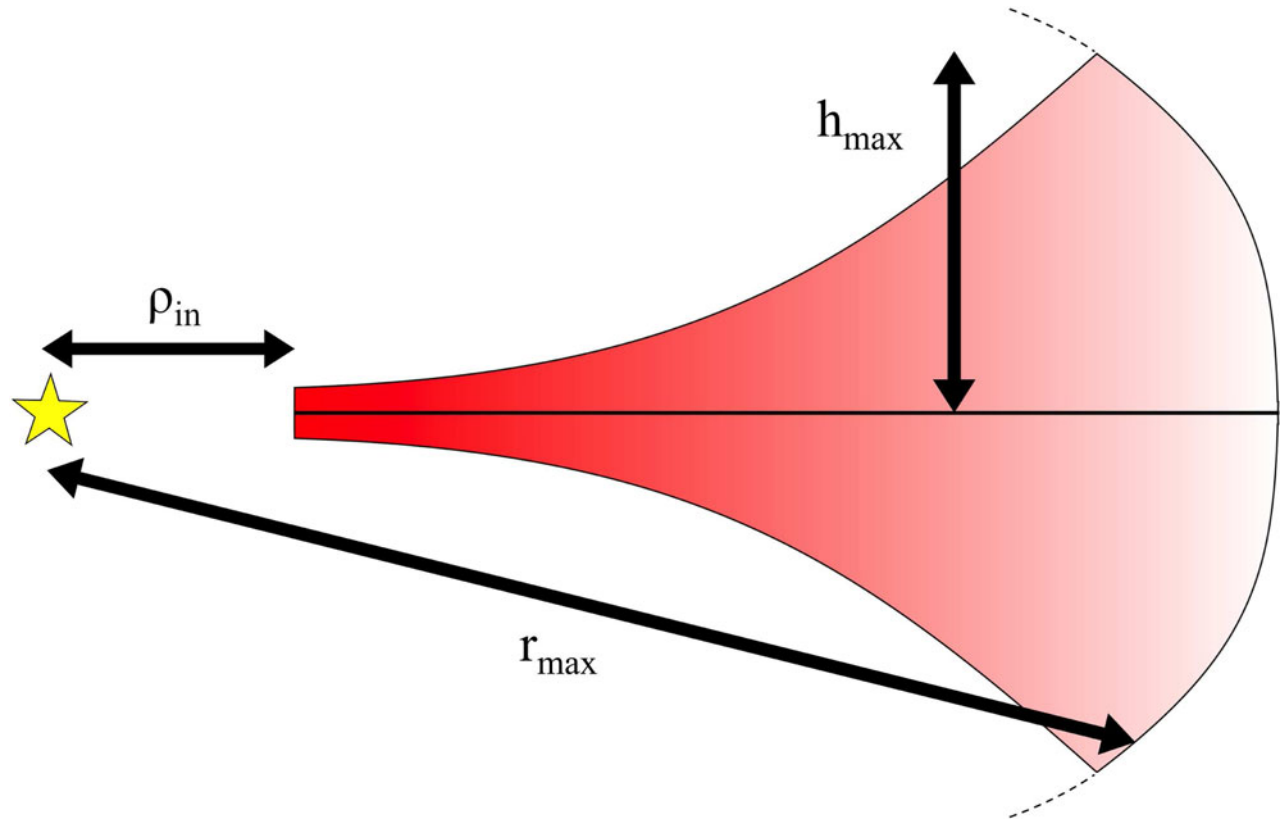

**Figure 3.** Sketch showing the different parameters that define the structure of the disc. $\rho_{in}$ is the inner cylindrical radius of the disc while $r_{max}$ is its outer spherical radius, shown by the dotted lines. $h_{max}$ is taken from the mid-plane of the disc.

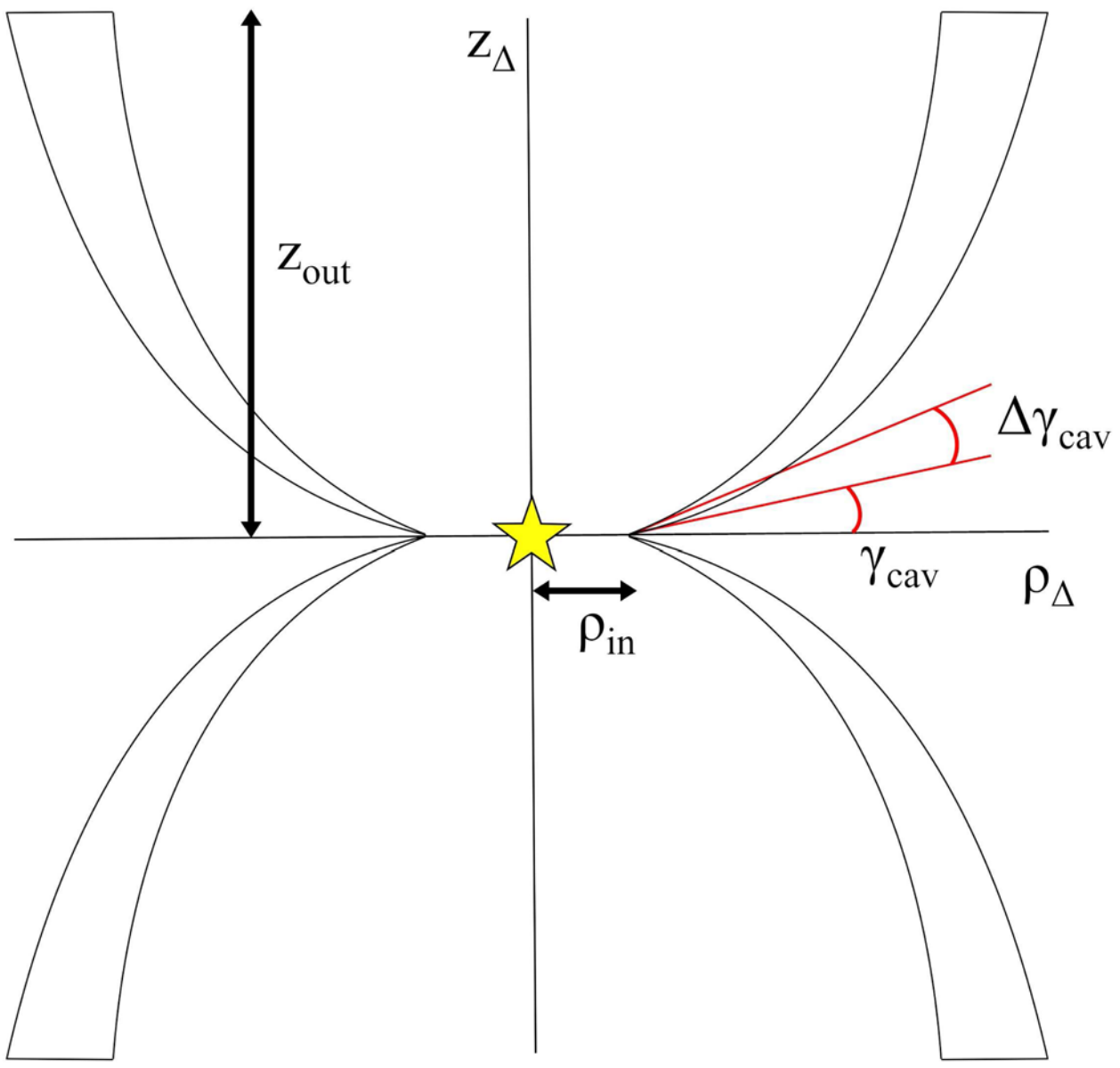

**Figure 4.** Sketch showing the different parameters that define the structure of the outflow. $\rho_{in}$ is the inner cylindrical radius of the outflow while $z_{out}$ is the maximum height that the outflow can reach. $\rho_\Delta$ and $z_\Delta$ define the cylindrical axes of the outflow. $\Delta\gamma_{cav}$ defines the width of the cavity walls.

The spherical source grid generation is rather simple since we only need to consider the inner $r_{in}^{sphere}$ and outer $r_{out}^{sphere}$ radii of the sphere. These radii will be the boundaries in which the grid points will be generated following the previous description.

The case of the disc grid generation is slightly different since we also have to consider the inner cylindrical and outer spherical radii ($\rho_{in}$ and $r_{max}$) and the maximum height ($h_{max}$) of the disc (see Fig. 3 and Section 3.2). This results in a different distribution of points compared to the spherical source. Indeed, grid points are generated over the entire volume of the cube defining the size of the modelled region, whereas the disc is a flattened structure, resulting in many points being generated outside of the edges. The Lloyd algorithm will then reject more points outside of the grid and to avoid this effect the number of iterations is limited to five and 5 per cent more points are added in this case. The middle panel of Fig. 2 displays a simulation of a disc grid generation by setting a total number of 5 000 points, an inner and outer radius of 1 arcsec and 10 arcsec, respectively. The maximum height of the disc is set to 2 arcsec. One can note from this figure that the Lloyd algorithm moves many points according to the different shape of the distribution before and after the smoothing process.

The outflow model is always considered attached either to a spherical source or a disc and cannot be created alone. None the less, for the outflow the point generation is performed as it is done for the disc (see the bottom panel of Fig. 2). In the outflow case, we consider the inner cylindrical radius $\rho_{in}$ and the maximum height $z_{out}$ of the outflow (see Fig. 4). The cavity angle $\tan\gamma_{cav} = z_{outflow}/\rho_{outflow}$ is defined as the angle between the points and the mid-plane (see Fig. 4). Each point of the grid is generated between $\gamma_{cav}$ and $\gamma_{cav} + \Delta\gamma$ where $\Delta\gamma$ is set by the user and defines the width of the cavity walls (see Section 3.3 for more details about the cavity walls). The smoothing process is here also limited to five iterations to avoid a great number of points to be moved outside of the model and 5 per cent more points are added as in the disc generation process. The bottom panel of Fig. 2 displays a simulation of an outflow cavity grid generation by setting a total number of 5 000 points, an inner and outer radius of 1 arcsec and 15 arcsec, respectively, with $a_{\rm outflow} = 150$ arcsec, $b_{\rm outflow} = 15$ arcsec and $\Delta\gamma = 5°$. $a_{\rm outflow}$ and $b_{\rm outflow}$ are respectively the major and minor axes of the outflow, assimilated to be an ellipse (see Section 3.3). The shape and the clear limit of the maximum height $z_{\rm out} = 15''$of the outflow can be identified in this plot. Visser et al. (2012) have used the same kind of outflow cavity wall gridding process.

It is also important to note that in the case where multiple structures are modelled at the same time, the total number of points will be equally distributed for each structure. For instance, for a total number of points of 15 000 and if one wants to model one spherical source, one disc and one outflow, GASS will distribute 5000 points for each structure.

## 3 CREATION OF THE PHYSICAL MODEL

Once the grid is built according to the different structures, GASS will determine which points belong to which structure. Since the grid points have been moved by the Lloyd's algorithm, some may not belong to any structure anymore but they can still be part of the computational domain. These points are important because they ensure the good continuity of the physical model between the structures and the edge of the grid. These points have a background density of 1 cm$^{-3}$ and gas and dust temperature of 2.73 K whereas all others parameters are set to zero.

### 3.1 Spherical sources generation

Each created Voronoï cell will be defined by a constant value for each of the physical parameters we consider: gas temperature, dust temperature, $H_2$ density, molecular abundance, velocity field and Doppler parameter. The given value for each cell is determined through an interpolation of the physical profile defined as input for each of these parameters. Since the grid is distributed randomly, each one of the Voronoï cells is situated at a different radius, thus it is uniquely defined by its physical properties. Both the density and temperature profiles can be defined with a variable number of regions $N_{\rm temp}$ and $N_{\rm dens}$, respectively, as a function of the radial distance from the central object by the following.

(i) Multiple power-law profiles: for each temperature region $i$, a power-law coefficient $\alpha_i$ and a temperature $T_{{\rm env},\,i}$ value is set for a given radius $r_{0,i}^{\rm temp}$, and for each density region $j$ a power-law coefficient $\beta_j$ and a density $n({\rm H}_2)_{{\rm env},\,j}$ value are set for a given radius $r_{0,j}^{\rm dens}$. The total physical profile is then defined as

$$T_{\rm env} = \sum_i^{N_{\rm temp}} T_{{\rm env},i} \left(\frac{r_i}{r_{0,i}^{\rm temp}}\right)^{\alpha_i}, \quad (1)$$

$$n({\rm H}_2)_{\rm env} = \sum_j^{N_{\rm dens}} n({\rm H}_2)_{{\rm env},j} \left(\frac{r_j}{r_{0,j}^{\rm dens}}\right)^{\beta_j}. \quad (2)$$

where $r_i$ and $r_j$ are all the radii that define the regions $i$ and $j$, respectively. Thus, the final temperature or density value of a region becomes the first temperature or density value of the following regions, ensuring the continuity of the physical profile.

(ii) Multiple temperature or density steps: for each region, a temperature or a density is set for a given radius. The code interpolates the temperature linearly and the density logarithmically between the two points. As above, different radii can be set for the temperature and the density profile.

The abundance profile can be defined in multiple regions $N_{\rm abund}$ as a function of the radius or the temperature (once the temperature profile is defined). This option can be used, for instance, to describe the freeze out or the desorption of a specific molecular species at a given temperature (e.g. CO at ∼27 K). The interpolation process is the same as for the temperature or the density profile but the user can choose a constant or a logarithmic variation of the abundance within the region. Each abundance value in each region can be gridded to calculate multiple models at the same time.

In the case that more than one structure (spherical source or disc) is used, the problem is to take into account their different contribution over the entire grid. This is only done for the spherical sources whereas outflows and discs impose their own physical properties in the region where they are defined, without considering the presence of the spherical sources (see Section 3.3). In any case, the outflow structure always prevails over other structures. For each point of the grid, the density of each spherical source is added following the equation:

$$n_{\rm cell} = \sum_i^N n_i, \quad (3)$$

where $n_{\rm cell}$ is the total $H_2$ density of the cell, $n_i$ is the $H_2$ density contribution of the spherical structure $i$ and $N$ is the number of different spherical structures. For the gas temperature, we consider the ideal gas law equation of state $P = nk_BT$, where $P$ is the pressure of the gas, $n$ is the number density, $k_B$ is the Boltzmann constant and $T$ is the absolute temperature. Since we consider a polyatomic species, each cell contains a total energy of

$$U_{\rm cell} = \frac{5}{2} k_B T_{\rm cell} n_{\rm cell} = \sum_i^N \frac{5}{2} k_B T_i n_i, \quad (4)$$

where $U_{\rm cell}$ is the internal energy of the cell, $T_{\rm cell}$ is the total temperature and $n_{\rm cell}$ is the total number density. The previous equation combined with equation (3) leads to

$$T_{\rm cell} = \frac{1}{\sum_i^N n_i} \times \sum_i^N T_i n_i. \quad (5)$$

The dust temperature can be defined separately but it can also be considered to be equal to the gas temperature.

The molecular abundance $[X]$ can be defined as

$$[X] = \frac{n_X}{n}, \quad (6)$$

where $n_X$ is the number density of the species. Considering the contribution of all the spherical sources, the abundance in each cell is

$$[X_{\rm cell}] = \frac{n_{X,{\rm cell}}}{n_{\rm cell}} = \frac{1}{\sum_i^N n_i} \times \sum_i^N [X]_i n_i, \quad (7)$$

where $[X_{\rm cell}]$ is the total molecular abundance in the cell and $n_{X,\,{\rm cell}}$ is the total number density of the species.

To calculate the radiative transfer one needs to define the total Doppler broadening, often called the $b$-Doppler parameter:

$$b = v_{\rm D} = \sqrt{v_{\rm th}^2 + v_{\rm turb}^2}, \quad (8)$$

where $v_{\rm turb}$ is the (micro-)turbulence velocity that operates on length-scales shorter than the photon mean free path. The thermal velocity, $v_{\rm th}$, is the random motion of molecules due to the kinetic temperature of the gas:

$$v_{\rm th} = \sqrt{\frac{2\,k_B\,T}{\mu\,m_{\rm H}}}, \tag{9}$$

Considering a Gaussian profile, the FWHM due to the Doppler broadening is

$$\mathrm{FWHM_D} = 2\sqrt{\ln(2)} \times v_{\rm D}, \tag{10}$$

where $\mathrm{FWHM_D}$ is the spectral line full width at half-maximum. Equation (10) can also be written as

$$b = \frac{1}{2\sqrt{\ln(2)}} \times \mathrm{FWHM_D} = 0.60 \times \mathrm{FWHM_D}. \tag{11}$$

In the current GASS version, the $b$-Doppler parameter can be considered either constant throughout the grid or variable as a function of the grid radius. Its value is defined by the user in the interface.

The velocity field of the spherical source is determined by adding the different projections on the Cartesian coordinates ($X$, $Y$, $Z$) of each velocity field induced by each structure included in the model. The intensity of the velocity field for each point of the grid for the spherical model is defined as an infall model by

$$V_{\rm inf}(r) = \sqrt{\frac{2\,G\,M_\star}{r}}, \tag{12}$$

where $v_{\rm inf}$ is the infall velocity, $G$ is the gravitational constant, $M_\star$ is the mass of the central object and $r$ is the distance from the central object. The projections on each Cartesian coordinates are calculated by

$$r = \sqrt{(X - X_\star)^2 + (Y - Y_\star)^2 + (Z - Z_\star)^2}, \tag{13}$$

$$\theta = \arctan\left(\frac{\sqrt{(X - X_\star)^2 + (Y - Y_\star)^2}}{Z - Z_\star}\right), \tag{14}$$

$$\phi = \arctan\left(\frac{(Y - Y_\star)}{(X - X_\star)}\right), \tag{15}$$

where $X_\star$, $Y_\star$ and $Z_\star$ are the coordinates of the central object. This leads to the following equations for the projection of the velocity vector:

$$\begin{cases} \boldsymbol{V}_x = -r\sin\theta\cos\phi\,\boldsymbol{e}_x, \\ \boldsymbol{V}_y = -r\sin\theta\sin\phi\,\boldsymbol{e}_y, \\ \boldsymbol{V}_z = -r\cos\theta\,\boldsymbol{e}_z, \end{cases} \tag{16}$$

where $\boldsymbol{e}_x$, $\boldsymbol{e}_y$ and $\boldsymbol{e}_z$ are the unit vectors. The velocity vector for each cell is then defined from the velocity of each structure $i$ by

$$\begin{cases} \boldsymbol{V}_{x,\rm cell} = \sum_i^N \boldsymbol{V}_{x,i}, \\ \boldsymbol{V}_{y,\rm cell} = \sum_i^N \boldsymbol{V}_{y,i}, \\ \boldsymbol{V}_{z,\rm cell} = \sum_i^N \boldsymbol{V}_{z,i}. \end{cases} \tag{17}$$

If the spherical structure is too complicated to be described through the interface, it is also possible, for each spherical model, to feed GASS with a file containing information about any of the following parameters as a function of the radius: density, gas and dust temperatures, molecular abundance and velocity field intensity.

### 3.2 Disc generation

The main difference between the spherical source model and the disc model is the number of symmetries. In the spherical case, every physical parameter can be defined as a function of the radius. In the disc model, there is only one symmetry around the rotational axis of the disc. The physical properties are therefore defined as a function of both the radius $\rho$ and the height $z$. One must take care of the difference between $r$ and $\rho$: $r$ is the spherical radial distance and $\rho$ is the cylindrical radial distance. The link between the two radii is given by $r = \sqrt{\rho^2 + z^2}$. As said above, we have to consider the inner and outer radius ($\rho_{\rm in}$ and $r_{\rm max}$) and the maximum height ($h_{\rm max}$) of a disc. Every point of the grid must be included between these values and they must also be at a smaller height than the pressure scaleheight $h_0$, which is defined as follow (Brinch & Hogerheijde 2010):

$$h_0 = \sqrt{\frac{2\,T_{\rm mid}\,k_B\,\rho^3}{G\,M_\star\,m_{\rm H}}}, \tag{18}$$

with $T_{\rm mid}$ the mid-plane temperature of the disc, $k_B$ the Boltzmann constant, $G$ the gravitational constant, $M_\star$ the mass of the central object and $m_{\rm H}$ the hydrogen atom mass. To calculate this value, we need to define the mid-plane temperature gradient across the disc. This temperature can be defined by a power law (Williams & Best 2014):

$$T_{\rm mid} = T_{\rm mid,0}\left(\frac{\rho}{\rho_{\rm in}}\right)^\gamma, \tag{19}$$

where $T_{\rm mid,\,0}$ is the mid-plane temperature at the radius $\rho_{\rm in}$. The atmosphere temperature profile of a disc is also defined as the temperature profile at a specific height $z = 4\,h_0$ of the disc (see Williams & Best 2014). The atmosphere temperature is set the same way the mid-plane temperature is defined, by a power law:

$$T_{\rm atm} = T_{\rm atm,0}\left(\frac{\rho}{\rho_{\rm in}}\right)^\gamma, \tag{20}$$

with $T_{\rm atm,\,0}$ the atmosphere temperature at the radius $\rho_{\rm in}$. All the characteristic values ($T_{\rm mid,\,0}$, $T_{\rm atm,\,0}$, $\rho_{\rm in}$ and $\gamma$) can be set in the GASS user interface and plots are made to facilitate the visualization of the results. The resulting temperature $T(\rho, z)$ in each cell of the disc as a function of $\rho$ and $z$ is (Williams & Best 2014)

$$T(\rho, z) = \begin{cases} T_{\rm mid} + (T_{\rm atm} - T_{\rm mid})\left[\sin\left(\frac{\pi\,z}{4\,h_0}\right)\right]^4 & \text{if } z < 4\,h_0 \\ T_{\rm atm} & \text{if } z \geq 4\,h_0 \end{cases}. \tag{21}$$

The density distribution is based on the profile defined in e.g. Chiang & Goldreich (1997) or Dullemond & Dominik (2004) by

$$n_{\rm H_2}(\rho, z) = n_0\left(\frac{\rho}{\rho_{\rm in}}\right)^\delta \exp\left[-\left(\frac{z}{h_0}\right)^2\right], \tag{22}$$

with $n_0$ the $H_2$ density at $\rho_0$. As for the temperature profile, $n_0$ and $\delta$ can be set in the interface. The way GASS deals with the abundance profile generation is the same as for the envelope. The abundance profile can be defined as a function of the cylindrical radial distance $\rho$ or as a function of the total temperature profile $T(\rho, z)$ of the disc. Thus, according to the choice of these two options, the abundance profile will depend on the disc height $z$.

The cylindrical axis $\Delta$ of the disc model can be rotated as a function of two angles, $\Theta$ and $\Phi$, thanks to the rotation matrix

$$R_{(\Theta,\Phi)} = \begin{pmatrix} \cos(\Phi) & 0 & -\sin(\Phi) \\ \sin(\Theta)\sin(\Phi) & \cos(\Theta) & \sin(\Theta)\cos(\Phi) \\ \cos(\Theta)\sin(\Phi) & -\sin(\Theta) & \cos(\Theta)\cos(\Phi) \end{pmatrix}, \tag{23}$$

where $\Theta$ is the angle between the axis $\Delta$ and the $z$-axis and $\Phi$ is the angle between the $x$-axis and projection of the axis $\Delta$ in the (X, Y) plane (see Fig. 5).

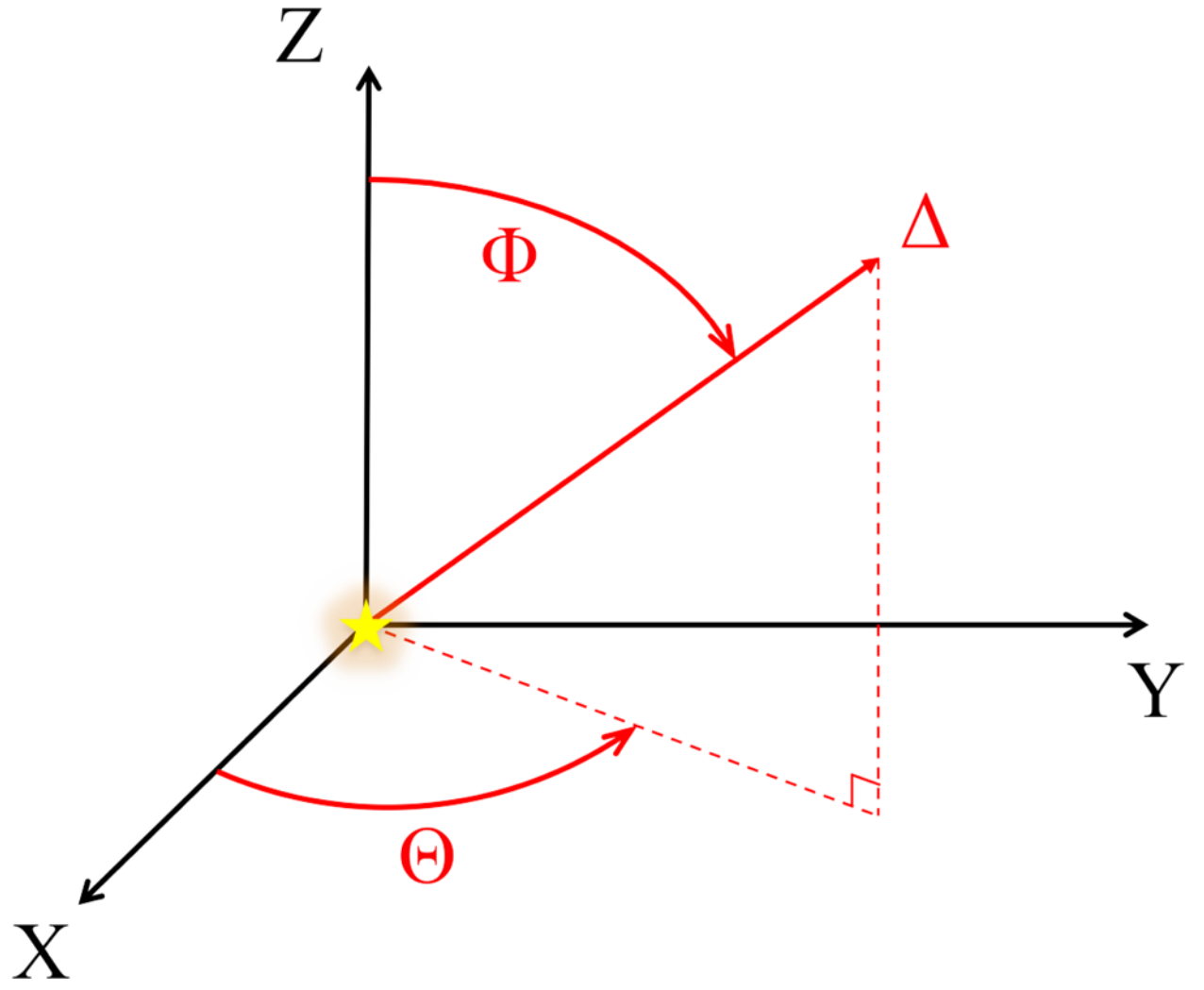

**Figure 5.** Sketch showing the different angles that are used by the code to rotate the disc and the outflow models. Δ represents the axis on which the model is constructed.

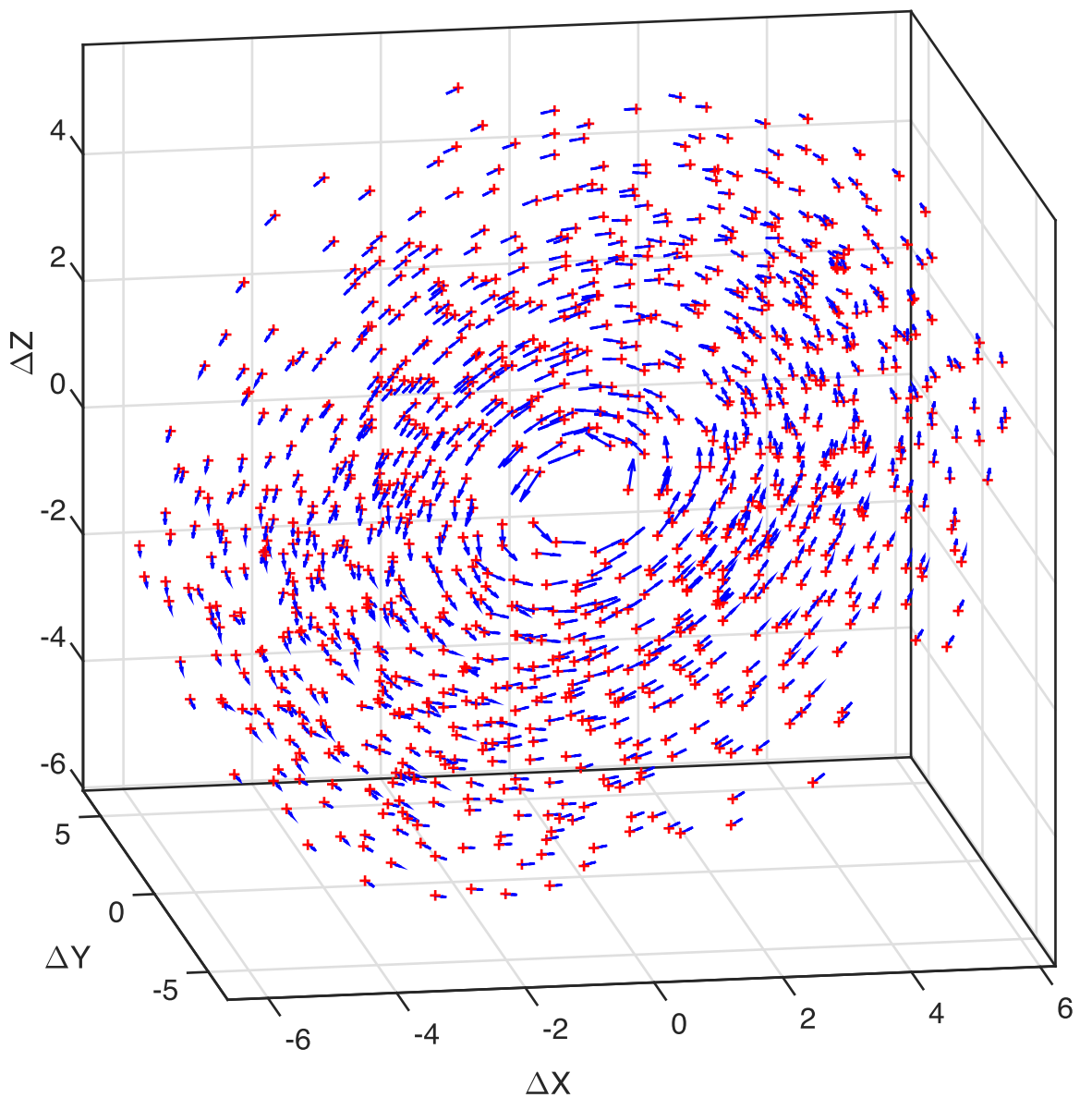

**Figure 6.** Grid points distribution (red crosses) of the disc model described in Section 3.2. The inner radius $\rho_{\rm in}$ is clearly visible as well as the two rotation angles Θ and Φ. The velocity vectors are plotted in blue and show the Keplerian rotation of the disc.

Fig. 6 shows a disc located at 300 pc generated with 10 000 points with $\rho_{\rm in} = 0.7$ arcsec, $r_{\rm max} = 7$ arcsec, $\Theta = 45^\circ$ and $\Phi = 45^\circ$. In this figure is also displayed the velocity field, supposed to follow the Keplerian rotation with the equation

$$V_{\rm rot}(\rho, z) = \sqrt{\frac{G\,M_\star}{(\rho^2 + z^2)^{1/2}}}. \quad (24)$$

Fig. 7 shows the resulting positions of the grid points in cylindrical coordinates plotted with the pressure scaleheight $h_0$. This plot verifies that the program correctly rejects as disc points any point higher than the pressure scaleheight.

GASS also computes the total gas mass of the disc after generating the grid. An accurate mass calculation would need to consider the volume of each Voronoï cell but this is not trivial, so we chose to consider the volume of the Delaunay triangulation associated with the Voronoï diagram instead. For each tetrahedron of the Delaunay triangulation, we consider a constant density determined from the input density profile. Since LIME also considers a constant density in a given Voronoï cell, we are close to the mass seen by LIME even if we are calculating the mass based on the Delaunay triangulation instead of the Voronoï diagram. With $v_i$ the volume of a given Delaunay tetrahedron $i$ (see Appendix A) of density $n(H_2)_i$, the mass is defined by

$$M_{\rm disc}^{\rm gas} = \frac{m_{H_2}}{M_\odot} \times \sum_i n(H_2)_i\, V_i, \quad (25)$$

with $m_{H_2}$ the mass of $H_2$. Again, this mass is close, but not exactly, the one that LIME will see and therefore which the output image is based on. The mass estimated by GASS is only informative and helps the user checking if the mass of the disc is physically meaningful or not.

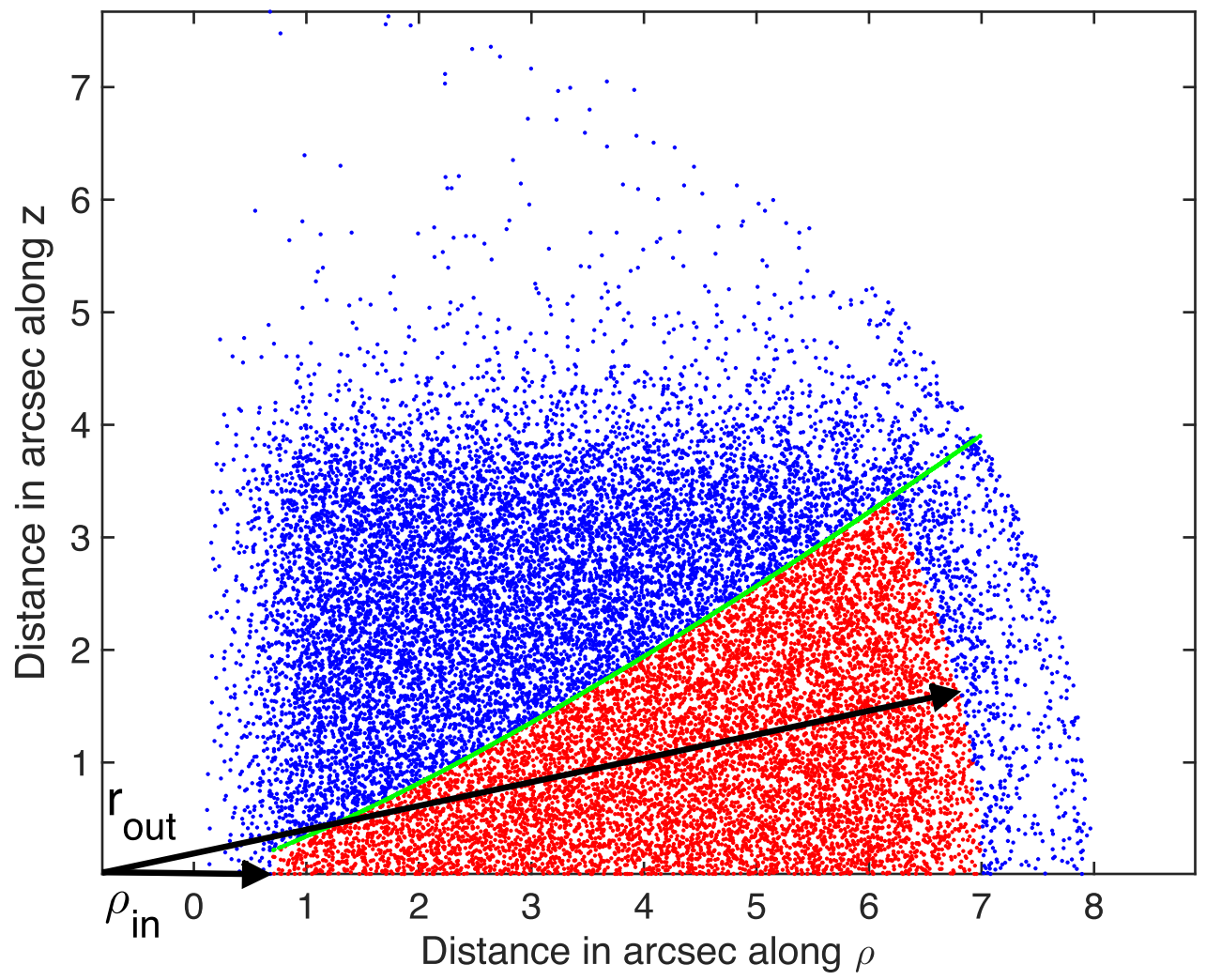

**Figure 7.** Disc profile in cylindrical coordinates ($\rho$, $z$). Disc grid points are shown in red whereas rejected points are in blue. These points are still in the computational domain but they have been rejected since they are above the specific pressure scaleheight $h_0$, plotted in green. The inner cylindrical and outer spherical radii ($\rho_{\rm in}$ and $r_{\rm out}$) can be identified as well.

As for the spherical structure, it is also possible, for each disc model, to feed GASS with a file containing pieces of information about any of the following parameters as a function of the radius and height: density, gas and dust temperatures, molecular abundance and velocity field intensity. For the gas temperature, it is possible to define either both $T_{\rm atm}$ and $T_{\rm mid}$ (as a function of the radius only) or the ‘final’ gas temperature (as a function of the radius and height).

### 3.3 Outflow generation

To date, one can consider three distinct types of outflows driven by jets or winds and a summary of their properties has been given in Arce et al. (2007).

(i) The jet bow-shock model with a highly collimated jet blowing the envelope away and creating a thin outflow shell (cavity walls) around the jet.

(ii) The wind-driven shell model, with a wide-angle radial wind and a thin shell interacting with the envelope.

(iii) The turbulent jet model, with Kelvin–Helmholtz instabilities along the jet/environment boundary leading to a turbulent viscous layer.

Arce & Sargent (2006) and Arce et al. (2007) have shown that in reality these different kinds of outflows are probing different outflow ages, thus probing different protostellar stages (Cantó, Raga & Williams 2008). Young outflows tend to be associated with the outflow type (i) with highly collimated jets and a faint wind, around Class 0 objects. As the protostar evolves, the loss of surrounding materials leads to a less dense environment around the jet. The outflow becomes wider and the wind stronger, now associated with the outflow type (ii), observed in Class I protostars. This trend can be used to estimate the age of a low-mass protostar (see Arce & Sargent 2006, especially their discussion section).

GASS can deal with both the jet bow-shock outflow model (iii) and the wind-driven shell model (i) by setting the appropriate value of the parameters $a_{\rm outflow}$, $b_{\rm outflow}$ and $\Delta\gamma$ (see Fig. 4).

As mentioned in Section 2, the outflow model is based on the mathematical definition given by Visser et al. (2012) and assimilated to an ellipse (or a part of an ellipse) centred on a central object with $a_{\rm outflow}$ and $b_{\rm outflow}$ the ellipse parameters. The outflow is modelled around an axis $\Delta$ (first superimposed to the $z$-axis) and the height of the outflow $z_\Delta$ is defined as a function of the cylindrical radial distance $\rho_\Delta$ from $\Delta$ by

$$z_\Delta = b_{\rm outflow}\sqrt{1-\left(\frac{\rho_\Delta}{a_{\rm outflow}}-1\right)^2}. \qquad (26)$$

The model is considered to be bipolar and symmetric with respect to the (X, Y) plane but the user can choose only one part of the outflow in the interface if needed. In GASS, outflows are always associated with a spherical structure or a protoplanetary disc and cannot be modelled alone. Thus, there is always a central object that defines the centre coordinates of the outflow. The axis $\Delta$ can be rotated in the model as a function of the two angles $\Theta$ and $\Phi$ thanks to the rotation matrix described in the previous section. The size of the outflow is limited by an inner radius $r_{in}$ and a maximum height $z_{\rm out}$ determined by the user in the interface. The code will then identify which points belong to the outflow structure, thus no points are added to the model.

In its region of influence, the outflow imposes its physical parameters as if its gas blows away the spherical envelope or the disc when it forms. The $H_2$ density $n_{H_2}$, the gas temperature $T_{\rm gas}$ and the abundance $[X]$ are defined using a power law as a function of the cylindrical radial distance $\rho$ from the central object:

$$\left\{n_{H_2}, T_{\rm gas}, [X]\right\} = \{n_0, T_0, [X]_0\}\left(\frac{\rho}{\rho_{\rm in}}\right)^{\{\epsilon,\zeta,\eta\}}, \qquad (27)$$

where $n_0$, $T_0$ and $[X]_0$ are, respectively, the density, temperature and abundance value at $\rho_{\rm in}$. $\epsilon$, $\zeta$ and $\eta$ are the power-law indices associated with each of this parameter, respectively. The velocity field is defined along the shape of the outflow with a constant value, taking into account that the velocity vector is always parallel to the $\Delta$ axis. Since the outflow gas is still subject to the gravitational field of the central object, the velocity vector of the outflow is added to the spherical model one. Fig. 8 shows the grid points that are assimilated to an outflow model located at 120 pc. The outflow parameters are: $\rho_{\rm in} = 1$ arcsec, $\rho_{\rm out} = 15$ arcsec, $a_{\rm outflow} = 150$ arcsec, $b_{\rm outflow} = 15$àrcsec, $\Theta = 45°$ and $\Phi = -45°$.

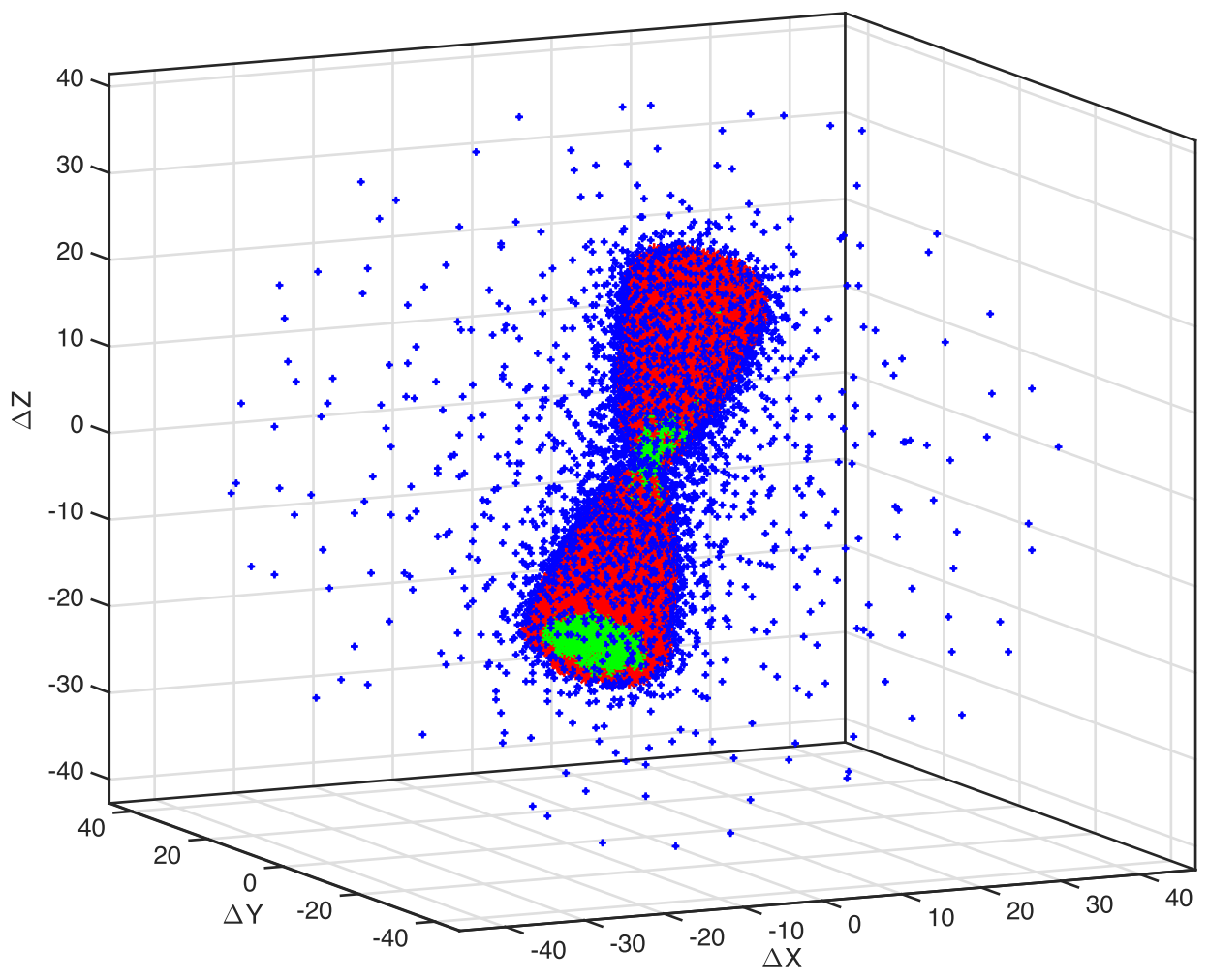

**Figure 8.** All the points have been generated by the outflow grid process. The red points have been identified among these points as the outflow cavity structure and now belongs to its model. All the other points have been rejected, the blue and green ones are, respectively, outside and inside the cavity walls of the outflow.

## 4 LIME OPTIONS IMPLEMENTED IN GASS

Several useful options of LIME normally defined in the input `model.c` file can be directly set in the GASS interface. GASS will then simply create the `model.c` file, taking into account all the LIME parameters set by the user. GASS produces a directory containing all the files required by LIME. The user just needs to launch LIME within this directory to obtain the output fits files. The user can directly describe the parameters of the output data cubes produced by LIME by giving the central frequency, the channel resolution, the bandwidth, the number of pixels per dimension, the pixel size and the name of the output fits files. The unit of these fits files can also be selected here, choosing between K, Jy pixel$^{-1}$ or S.I. units. Finally, the GASS interface also allows the user to select a dust opacity and a collision file (if needed). These latter are provided by data bases such as the LAMDA[5] data base (Schöier et al. 2005).

LIME does not necessarily calculate line emissions and it is possible to only choose to calculate the dust continuum emission. If so, the channel resolution, the bandwidth and the collision file are not required and the output fits is simply a continuum image.

GASS can also set the LTE mode of LIME. If it is the case, the population levels are directly calculated and no iterations are made, the output fits is then generated. This is very useful for molecules with no existing collisional rates or if non-LTE calculations are not required. However, one must be careful because in any case, the collision file (even if it does not contain collisional rates) is still required by LIME, since it is from this file that the code retrieves the spectroscopic parameters of the studied molecule, i.e. the Einstein coefficient $A_{\rm i,j}$, the frequency, or the upper energy level $E_{\rm up}$ of the transitions.

A useful option coded in GASS is to consider the ortho-to-para ratio of the $H_2$ molecule. Regarding other molecules, it is up to the user to correctly describe the ortho-to-para ratio of the studied molecule. Since frequencies of ortho and para species are not the same, their spectroscopic parameters are usually gathered in separated collision files. Thus, from the point of view of LIME, they are treated as two completely different molecules and two radiative transfer calculations are required to get the result of both ortho and para forms.

[5] Leiden Atomic and Molecular Database, http://home.strw.leidenuniv.nl/~moldata/

All available LIME options are not yet implemented in GASS (anti-alias, line blending, polarization,...) mainly because we need more time to include them into the code. Another reason is that some options have not yet been revised by the LIME development team since the release of the version 1.5, and they may cause issues or wrong results. It is the case of the line blending option for instance. As a consequence, we chose not to implement them yet in GASS.

LIME parameters need to be set correctly depending on the physical case treated by the user and a bad set of input parameters may lead to wrong results. An example of this effect is shown in Appendix B, where we discuss for instance the influence of the pixel size and the *anti-alias* option of LIME on the resulting image. For those who are not intimately familiar with LIME, one must take care of the following input properties that can be set in LIME, depending on the physical problem treated.

(i) If too few points are set in the model, the population density will not be calculated correctly and the spatial coverage of the model will not be smooth enough. This will result in visible and non-desired structure in the output image.

(ii) A pixel size much larger (or an *anti-alias* value too low, see Appendix B) than the mean scale of variation of the physical properties set in the model will lead to a wrong continuum calculation. Since the pixel is too large, it will not probe correctly the physics occurring in the region it covers and the pixel intensity will only reflect an average value of what is going on on a smaller scale.

Finally, one can note that if the size of the data cube is too large (large number of pixels and/or channels), LIME will take a long time to process it, leading to an excessively long calculation time while running a grid of models.

## 5 POST-TREATMENT ANALYSIS OPTIONS

To analyse hyper-spectral data cubes (generated by LIME for instance), GASS offers several functionalities to deal with the output fits files, depending on the observations in the hands of the user or depending on the processing one wants to perform with the models.

### 5.1 'Smoothing tool'

A 'smoothing tool' allows the user to average a certain number of model. Artefacts can appear in the data cube images, produced by LIME. This is mainly due to a lack of points in the outermost part of the grid, even if a smoothing process is already done through the Lloyd algorithm (see Section 2) in order to homogenize the distribution of points. Since each model is built with a different grid in GASS, averaging several runs reduces the artefacts due to the grid. Moreover, it allows us to reduce the initial number of points in the grid and the total execution time of a run in LIME depends a lot on this initial number of points. Therefore, it is faster and more efficient to run 10 models with 10 000 points at the same time and average them rather than doing a single run with 100 000 points. The smooth option implemented in GASS does this procedure automatically and creates a resulting smoothed fits file for each transition of a given model. We want to point out that an updated version of LIME coming soon will drastically reduce the appearance of artefacts in data cubes. An example of the averaging process is shown in Fig. 9 where the integrated intensity over all channels is plotted for one model only compared to the smoothed one.

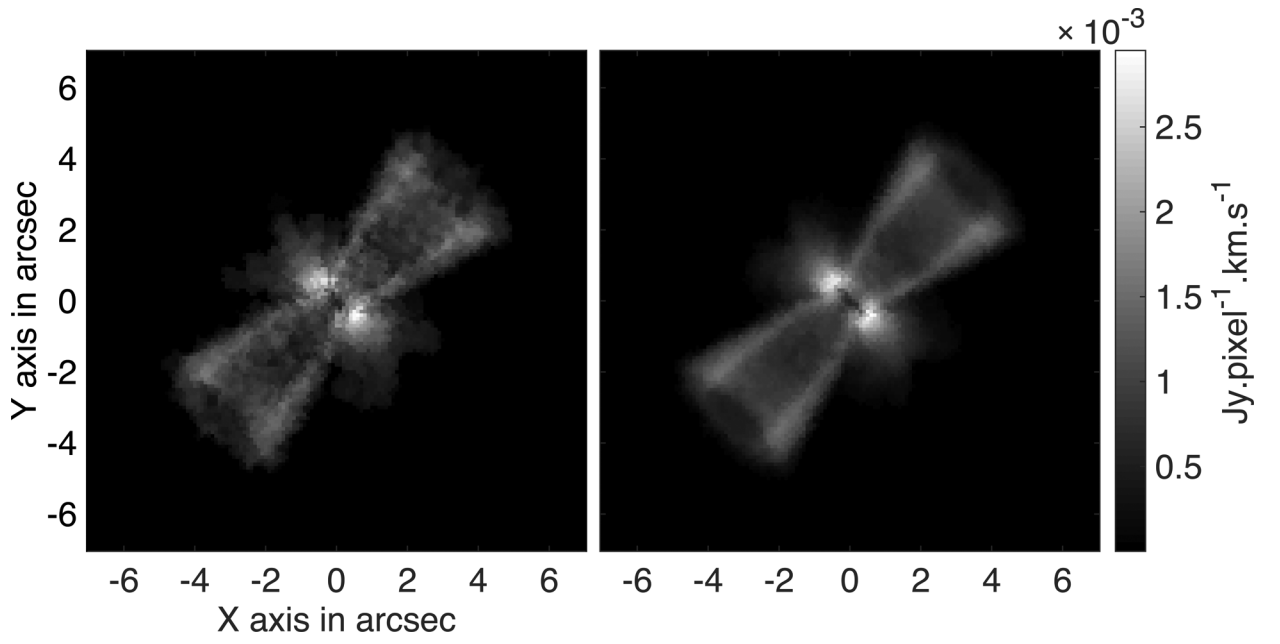

**Figure 9.** Integrated intensity map over all channels for one model only (left-hand panel) and 10 models averaged with the 'smoothing tool'.

### 5.2 Integrated maps

Integrated maps (moment 0, $M_0$, moment 1, $M_1$ and moment 2, $M_2$) of a given data cube can be calculated by the program, considering the following equations:

$$M_0 = \int I(v)\,\mathrm{d}v, \tag{28}$$

$$M_1 = \frac{\int I(v)\,v\,\mathrm{d}v}{\int I(v)\,\mathrm{d}v} = \frac{\int I(v)\,v\,\mathrm{d}v}{M_0}, \tag{29}$$

$$M_2 = \sqrt{\frac{\int I(v)\,(v - M_1)^2\,\mathrm{d}v}{M_0}}, \tag{30}$$

where $I$ is the pixel intensity and $v$ the velocity. $M_0$ represents the integrated intensity, $M_1$ the velocity field and $M_2$ the velocity dispersion field. The integral is calculated over the desired number of channels. The graphical interface allows the user to choose the channels over which one wants to calculate the moments and plot the results.

### 5.3 Plots of best modelled versus observed spectra

Another tool allows us to plot the resulting spectra of each transition in order to compare them to single-dish observations. The user gives in input a formatted file containing the information about the observations such as the name of the telescope (or the size of the antenna), the rms and a table gathering the frequencies and intensities for each spectra. GASS can directly read CASSIS[6] line files (`.lis` format) to perform this comparison. If several modelling directories exist in the current working directory, GASS reads all data cubes (fits files) produced by LIME in all these different modelling directories. Thanks to the graphical interface, the user can choose to display a specific data cube of a selected modelling. From this interface, it is possible to span the different channels of the data cube and the user can select the desired pointing position of the telescope with a cursor or directly choose the centre of the map.

Once the pointing position is set, GASS will produce a spectrum for each data cube by convolving the model with the different antenna beam sizes. We have taken these beam sizes from the different telescopes' respective websites and implemented them in GASS. For the IRAM-30m,[7] beam sizes are calculated following:

$$\theta_{\mathrm{IRAM-30m}}(\,\mathrm{arcsec}) = \frac{2460}{\nu}, \tag{31}$$

where $\nu$ is the frequency in GHz. We have derived the JCMT telescope beam size equation from a power-law fitting of beam values given at some frequencies on the JCMT website[8] (see Table 1):

$$\theta_{\mathrm{JCMT}}(\mathrm{arcsec}) = 801.6 \times \nu^{-0.6377} - 4.65, \tag{32}$$

where $\nu$ is the frequency in GHz.

[6] CASSIS is a software developed by IRAP-UPS/CNRS (http://cassis.irap.omp.eu).

**Table 1.** Beam sizes for the JCMT telescope (taken from the website) at given frequencies.

| Frequency (GHz) | HPBW (arcsec) |
|---|---|
| 150 | 28 |
| 230 | 21 |
| 345 | 14 |
| 690 | 8 |
| 870 | 6 |

**Table 2.** Beam sizes for the *Herschel*/HIFI telescope (taken from the HIFI beam release note of 2014 September) at given frequencies.

| Frequency $\nu_0$ (GHz) | HPBW $\theta_0$ ( arcsec) |
|---|---|
| 480 | 43.30 |
| 640 | 32.85 |
| 800 | 26.05 |
| 960 | 21.80 |
| 1120 | 19.50 |
| 1410 | 14.80 |
| 1910 | 11.10 |

For APEX, we directly took the equation on the telescope's website[9]:

$$\theta_{\mathrm{APEX}}(\mathrm{arcsec}) = 7.8 \times \left(\frac{800}{\nu}\right), \tag{33}$$

where $\nu$ is the frequency in GHz. Finally, for the *Herschel*/HIFI telescope, the HIFI beam release note of 2014 September[10] gives a general law to calculate beam sizes, depending on the HIFI band:

$$\theta_{\mathrm{HIFI}}(\mathrm{arcsec}) = \theta_0 \times \left(\frac{\nu_0}{\nu}\right), \tag{34}$$

with $\theta_0$ and $\nu_0$ gathered in Table 2.

A summary of beam sizes used by GASS for these telescopes as well as the frequency range available (with both PI and non-PI instruments) is plotted in Fig. 10. If the user is using any other telescope, GASS only needs the diameter $D$ (in m) of the dish to calculate the beam according to the theoretical equation:

$$\theta_B(\mathrm{arcsec}) = 1.22 \times \left(\frac{\lambda}{D}\right) \times \frac{3600 \times 180}{\pi} \simeq \frac{7.54 \times 10^4}{(\nu/\mathrm{GHz}) \times D}, \tag{35}$$

where $\lambda$ is the wavelength and $\nu$ is the frequency.

[7] http://www.iram.es/IRAMES/mainWiki/Iram30mEfficiencies
[8] http://www.eaobservatory.org/jcmt/instrumentation/heterodyne/
[9] http://www.apex-telescope.org/telescope/
[10] http://herschel.esac.esa.int/twiki/pub/Public/HifiCalibrationWeb/HifiBeamReleaseNote_Sep2014.pdf. The HIFI Beam: Release #1

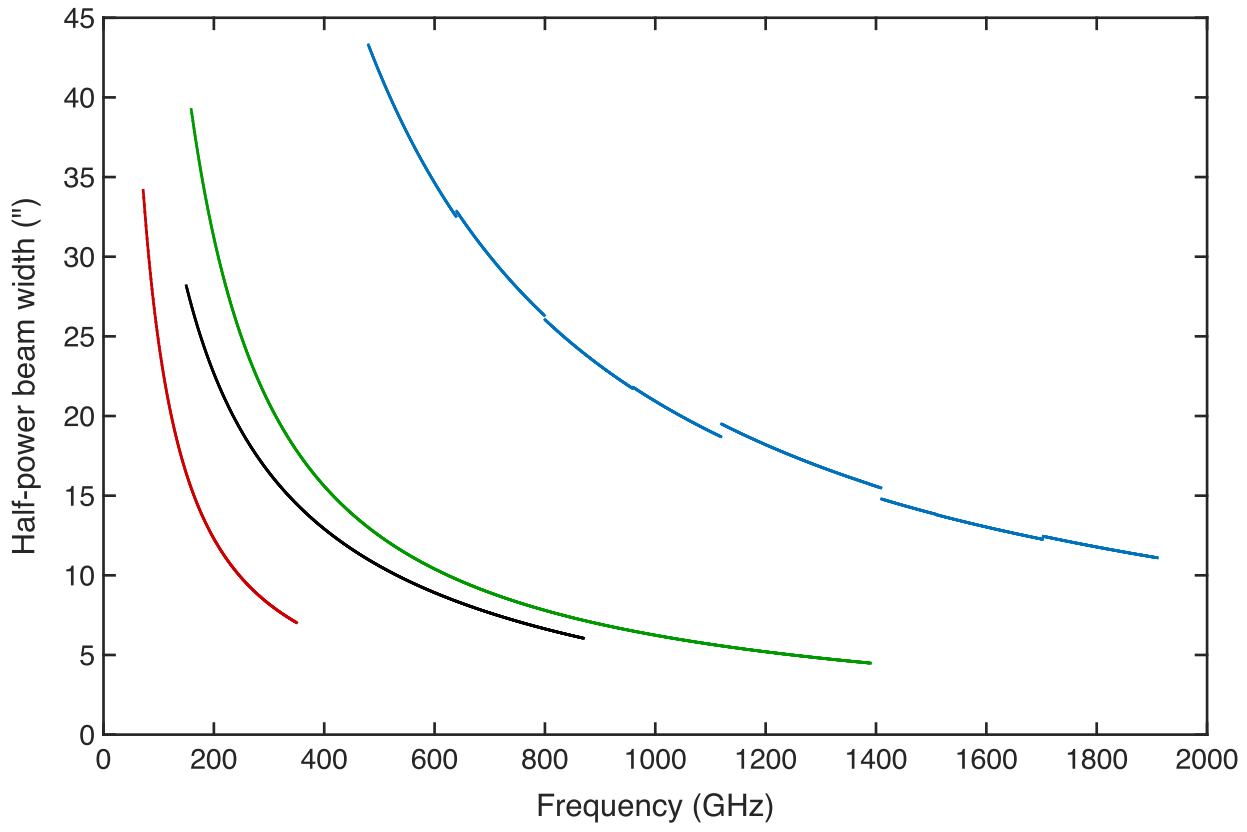

**Figure 10.** Half-power beam width (HPBW) as a function of the frequency for the IRAM-30m telescope (in red), JCMT (in black), APEX (in green) and *Herschel*/HIFI (in blue).

The calculated beam sizes can be wrong by a few per cents with respect to the real beam sizes of the telescope (which depends on the shape of the dish) but it is always an insignificant source of error in modelling results.

Once all the data cubes have been read and correctly convolved with GASS, the best-fitting model is calculated over all the models, following the standard $\chi^2$ minimization value for $N_{\mathrm{spec}}$ spectra $i$ of $N_i$ points, given by the equations (Lampton, Margon & Bowyer 1976):

$$\chi^2_{\mathrm{red}} = \frac{1}{\sum_i^{N_{\mathrm{spec}}} N_i} \sum_{i=1}^{N_{\mathrm{spec}}} \chi_i^2, \tag{36}$$

and

$$\chi_i^2 = \sum_{j=1}^{N_i} \frac{(I_{\mathrm{obs},ij} - I_{\mathrm{model},ij})^2}{\mathrm{rms}_i^2 + (\mathrm{cal}_i \times I_{\mathrm{obs},ij})^2}, \tag{37}$$

where $I_{\mathrm{obs},\,ij}$ and $I_{\mathrm{model},\,ij}$ are, respectively, the observed and the modelled intensity in the channel $j$ of the transition $i$, $\mathrm{rms}_i$ is the rms of the spectrum $i$, $\mathrm{cal}_i$ its calibration error. The method described in Lampton et al. (1976) uses a different formulae for the reduced $\chi^2$:

$$\chi^2_{\mathrm{red}} = \frac{1}{N_p - p} \sum_{i=1}^{N_{\mathrm{spec}}} \chi_i^2, \tag{38}$$

with $N_p$ equals $\sum_i^{N_{\mathrm{spec}}} N_i$ and $p$ is the degree of freedom of the minimization i.e. the number of adjustable parameters. It is difficult to trace this number since grid of models can be created by varying different parameters at the same time and not always the same ones. None the less, we have verified that $p \ll N_p$ thus $1/N_p \simeq 1/(N_p - p)$. For instance, fitting five transitions of 50 channels each by varying five different parameters leads to $N_p = 250$ and $p = 5$, which is much less than $N_p$.

This tool is useful to constrain results when combined with grid of models, for instance grid of abundance profiles. An example of this tool is given in Section 6.

### 5.4 Plots of best continuum model versus observations

The same analysis as for spectra can be done in GASS for continuum only models, using the same beam sizes. Rather than plotting spectra, this tool presents the continuum fluxes as a function of the frequency compared to the observations. None the less, the minimization is different and the $\chi^2$ is calculated for $N_{\text{cont}}$ continuum measurements with:

$$\chi^2_{\text{red}} = \frac{1}{N_{\text{cont}}} \sum_{i=1}^{N_{\text{cont}}} \chi_i^2, \tag{39}$$

and

$$\chi_i^2 = \frac{(F_{\text{obs},i} - F_{\text{model},i})^2}{(\text{cal}_i \times F_{\text{obs},i})^2}, \tag{40}$$

where $F_{\text{obs},\,i}$ and $F_{\text{model},\,i}$ are, respectively, the observed and the modelled intensity at the frequency $i$, and $\text{cal}_i$ the calibration error.

It is not possible at the moment to perform spectral energy distribution (SED) fitting due to the thermal emission of dust in GASS but we plan to incorporate it to the code by varying the different parameters of the following equation (Hildebrand 1983), especially $T_{\text{dust}}$ and $\beta$:

$$S_\nu = \Omega N \kappa_0 \left(\frac{\nu}{\nu_0}\right)^\beta B_\nu(T_{\text{dust}}). \tag{41}$$

This equation follows the Planck function, $B_\nu(T)$, calculated at the dust temperature $T_{\text{dust}}$. $N$ is the column density of dust, $\Omega$ is the solid angle of the observing beam and $\kappa_0\,(\nu/\nu_0)^\beta$ is the opacity of the emitting dust. This equation is valid at far-IR wavelengths ($\lambda \gtrsim 60\ \mu$m) and to use it the dust emission has to be optically thin, i.e. $\tau(\nu) \ll 1$. The optical depth $\tau$ of dust is calculated with:

$$\tau(\nu) = \int \kappa(\nu) \frac{\rho(\text{H}_2)}{\text{gas/dust}} \text{d}l, \tag{42}$$

where $\kappa(\nu) = \kappa_0\,(\nu/\nu_0)^\beta$, $\rho(\text{H}_2)$ is the volumetric mass density of $\text{H}_2$ and gas/dust is the gas-to-dust mass ratio ($\sim$100). The previous equation can be also written as

$$\tau(\nu) = \kappa_0 \left(\frac{\nu}{\nu_0}\right)^\beta \frac{m(\text{H}_2)}{\text{gas/dust}} \int n(\text{H}_2)\text{d}l, \tag{43}$$

$$= \kappa_0 \left(\frac{\nu}{\nu_0}\right)^\beta \frac{2m_p}{\text{gas/dust}} N(\text{H}_2). \tag{44}$$

with $m_p$ the mass of the proton and $N(\text{H}_2)$ the column density of $\text{H}_2$. Kelly et al. (2012, and references therein) have shown that there is a $T_{\text{dust}} - \beta$ degeneracy (producing an anti-correlation between the two) when doing $\chi^2$ minimization of SED fitting, leading to erroneous estimates of $T_{\text{dust}}$ and/or $\beta$ (see also Juvela & Ysard 2012). Different methods, including hierarchical Bayesian techniques (Kelly et al. 2012; Juvela et al. 2013), can be used to solve the problem but they can all introduce some bias in the analysis (Juvela et al. 2013) and care must be taken. We therefore delay the implementation of SED fitting to a future version of GASS.

### 5.5 Simulation of interferometric observations

In the case of interferometric observations, the analysis is more complicated. GASS possesses a tool that helps the comparison between the observations and the models.

A 2D Gaussian tool allows us to convolve the data cube with the observed beam of the interferometric data by giving the major and minor axes ($X_{\text{FWHM}}$ and $Y_{\text{FWHM}}$) of the beam and its position angle $\theta$. It is possible to directly compare the convolved predicted model with the observed data in a case where the $(u,\,v)$-coverage of the plane of sky is good and no flux is filtered out during the process. Therefore, this tool can be used to compare models with observations but it cannot be used to predict interferometric observations since it does not take into account the position and the number of antennas for instance. To do the convolution, first the beam area is calculated using

$$X_{\text{FWHM}}^{(\text{pix})} = \frac{X_{\text{FWHM}}^{(\text{arcsec})}}{\text{pix}_x} \quad \text{and} \quad Y_{\text{FWHM}}^{(\text{pix})} = \frac{Y_{\text{FWHM}}^{(\text{arcsec})}}{\text{pix}_y}, \tag{45}$$

where $X_{\text{FWHM}}^{(\text{pix})}$ and $Y_{\text{FWHM}}^{(\text{pix})}$ are respectively the $x$-axis and $y$-axis pixel size of the beam and $\text{pix}_x$, $\text{pix}_y$ are, respectively, the $x$-axis and $y$-axis pixel size in arcsec. The beam area in pixel/beam can be then calculated following:

$$\Theta_{\text{beam}} = \frac{\pi X_{\text{FWHM}}^{(\text{pix})} Y_{\text{FWHM}}^{(\text{pix})}}{4 \ln 2}. \tag{46}$$

Then the standard deviations $\sigma_x$ and $\sigma_y$ are calculated:

$$\sigma_x = \frac{X_{\text{FWHM}}^{(\text{pix})}}{2\sqrt{2 \ln 2}} \quad \text{and} \quad \sigma_y = \frac{Y_{\text{FWHM}}^{(\text{pix})}}{2\sqrt{2 \ln 2}}. \tag{47}$$

A 2D Gaussian can be defined as

$$f(x,\,y) = f_0 \exp\left[-\left(\frac{(x-x_0)^2}{2\sigma_x^2} + \frac{(y-y_0)^2}{2\sigma_y^2}\right)\right], \tag{48}$$

where $f_0$ is the amplitude and $(x_0,\,y_0)$ the centre. Generally, a 2D Gaussian can also be defined using:

$$f(x,\,y) = f_0 \exp\left[-\left(a(x-x_0)^2 + 2b(x-x_0)(y-y_0) + c(y-y_0)^2\right)\right], \tag{49}$$

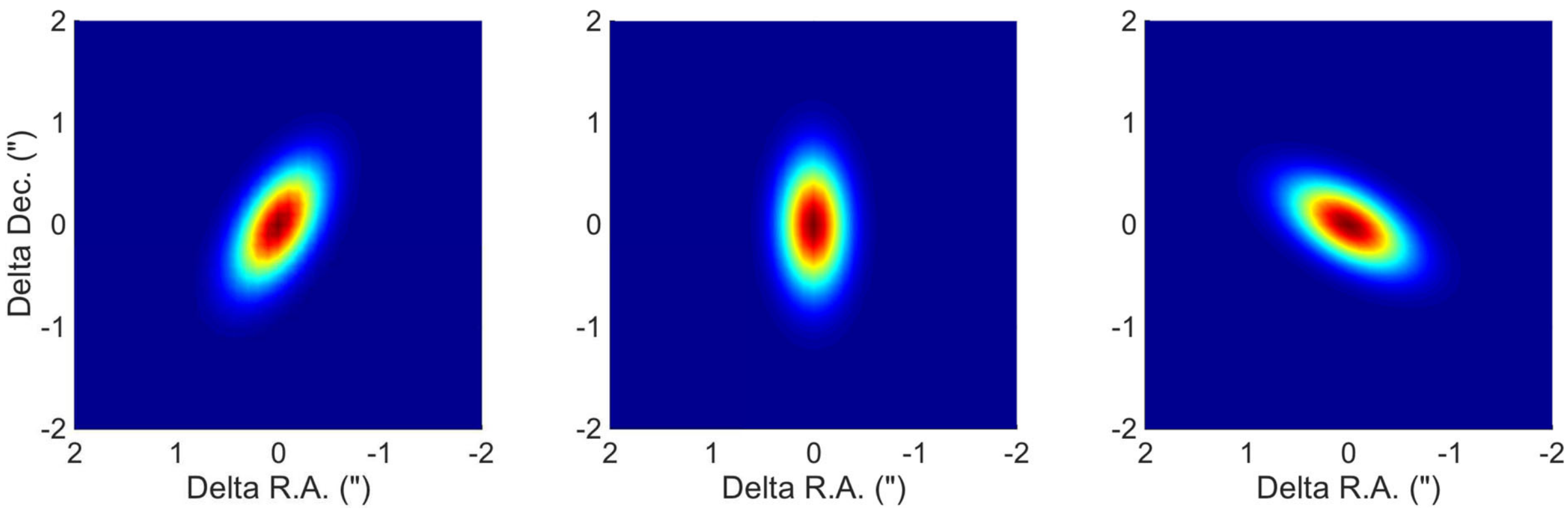

**Figure 11.** Final beam $F(x,\,y)$ shape for $X_{\text{FWHM}} = 0.5$ arcsec and $Y_{\text{FWHM}} = 1$ arcsec and $\theta = -30°$ (left-hand panel), $\theta = 0°$ (middle panel) and $\theta = 60°$ (right-hand panel) with respect to the $+y$-axis.

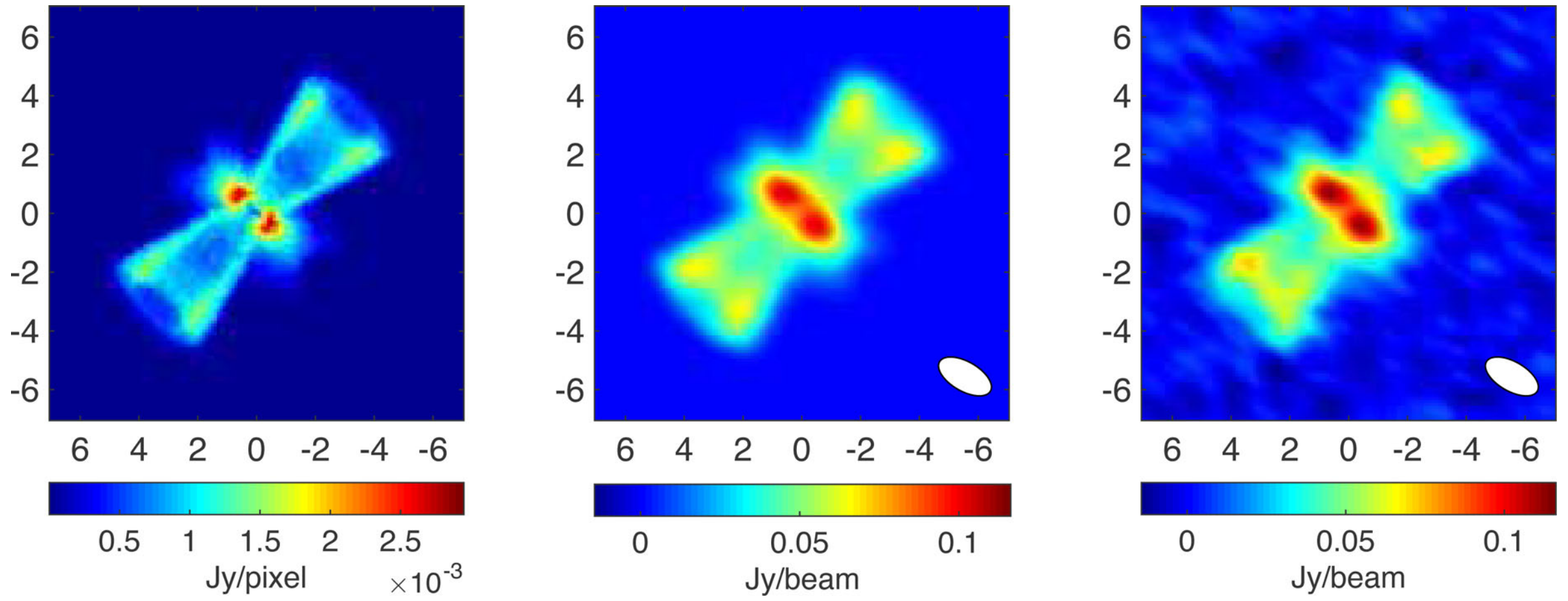

**Figure 12.** Left-hand panel: moment 0 map (in Jy pixel$^{-1}$ units) over all channels for the smooth model presented in the right-hand panel of Fig. 9. Middle panel: moment 0 map (in Jy beam$^{-1}$ units) resulting from the convolution between the data cube and a beam of $X_{\rm FWHM} = 0.5$arcsec, $Y_{\rm FWHM} = 1$arcsec, $\theta = 60°$. The full ellipse in the right corner shows the shape of the beam. Right-hand panel: same as the middle panel but with a white Gaussian noise of rms = 0.06 Jy beam$^{-1}$ per channel.

**Table 3.** Physical properties of the model set for the first example.

| Physical properties | Value |
|---|---|
| General properties | |
| Grid min, max radius | 0.1 arcsec, 7 arcsec |
| N° points | 10 000 |
| Distance | 300 pc |
| b-Doppler | 200 m s$^{-1}$ |
| $V_{\rm LSR}$ | 0 km s$^{-1}$ |
| Central object mass | 3 $M_\odot$ |
| Envelope properties | |
| $T_{\rm env,\,max}$ | 200 K ($\alpha = -0.5$) |
| $n(H_2)_{\rm env,\,max}$ | $1\times 10^9$ cm$^{-3}$ ($\beta = -1.5$) |
| $X_{\rm in}(^{13}CO)$, $X_{\rm out}(^{13}CO)$ | $1.5 \times 10^{-13}$, $7.5 \times 10^{-11}$ |
| Protoplanetary disc properties | |
| $\rho_{\rm in}$, $r_{\rm max}$, $h_{\rm max}$ | 0.7 arcsec, 4.5 arcsec, 5 arcsec |
| $\Theta$, $\Phi$ | (0, 45, 90)°, (0, 45, 90)° |
| $T_{\rm atm,\,in}$, $T_{\rm mid,\,in}$ | 500 K, 50 K ($\gamma = -0.5$) |
| $n(H_2)_{\rm disc,\,in}$ | $6.5\times 10^7$ cm$^{-3}$ ($\delta = -1.0$) |
| $X_{\rm in}(^{13}CO)$ for $T < 27$ K | $1.5 \times 10^{-6}$ |
| $X_{\rm out}(^{13}CO)$ for $T > 27$ K | $1.5 \times 10^{-17}$ |
| Outflow properties | |
| $a_{\rm outflow}$, $b_{\rm outflow}$, $\rho_{\rm in}$, $z_{\rm out}$ | 150 arcsec, 8 arcsec, 0.5 arcsec, 5 arcsec |
| $v_{\rm outflow}$ | 10 km s$^{-1}$ |
| $\Theta$, $\Phi$, $\Delta\gamma$ | (0, 45, 90)°, (0, 45, 90)°, 15° |
| $T_{\rm outflow}$ | 100 K |
| $n(H_2)_{\rm outflow}$ | $5 \times 10^6$ cm$^{-3}$ |
| $X(^{13}CO)$ | $1.5 \times 10^{-8}$ |

with:

$$a = \frac{\cos^2\theta}{2\sigma_x^2} + \frac{\sin^2\theta}{2\sigma_y^2}, \tag{50}$$

$$b = \frac{\sin(2\theta)}{4\sigma_x^2} - \frac{\sin(2\theta)}{4\sigma_y^2}, \tag{51}$$

$$c = \frac{\sin^2\theta}{2\sigma_x^2} + \frac{\cos^2\theta}{2\sigma_y^2}. \tag{52}$$

The signs in the $b$ coefficient determine the rotation of the Gaussian, defined as clockwise from the +y-axis here. For a counter-clockwise rotation, one needs to invert the signs in $b$. An example of the clockwise rotation is shown in Fig. 11 for several position angles $\theta$.

This 2D Gaussian is normalized using the volume $v$ under the Gaussian:

$$V = \int_{-\infty}^{+\infty}\int_{-\infty}^{+\infty} f(x, y)\,{\rm d}x{\rm d}y = 2\pi f_0\sigma_x\sigma_y. \tag{53}$$

In GASS, $f_0 = 1$ and the final beam function $F(x, y)$ used to be convolved with the predicted model is

$$F(x, y) = \frac{f(x, y)}{V}. \tag{54}$$

It is also possible to add a white Gaussian noise to the data cube before the convolution to reproduce the observed *rms*. The input data cube must be in Jy pixel$^{-1}$ and GASS can write an output fits file with the final results in Jy beam$^{-1}$ using the beam area $\Theta_{\rm beam}$ to perform the conversion from Jy pixel$^{-1}$ to Jy beam$^{-1}$. An example of this tool is shown in Fig. 12 where the input fits file is the smoothed data cube shown in Fig. 9. The beam is defined with $X_{\rm FWHM} = 0.5$ arcsec, $Y_{\rm FWHM} = 1$arcsec, $\theta = 60°$ (see right-hand panel of Fig. 11) for a beam area $\Theta_{\rm beam} \simeq 56.65$ pixel beam$^{-1}$.

It is possible with GASS to create output fits files with the convolved data cube with the beam shape written in the header. These files can be read with data cube analysis packages (CASA, GILDAS, DS9, etc.) to proceed with any further analysis. For instance, with the Common Astronomy Software Applications package (CASA), one can perform simulations of observations using directly the output hyper-spectral cube created by LIME, with the help of the *simobserve* and *simanalyze* CASA tasks.

## 6 EXAMPLES

In this section, we will show different examples of the GASS capabilities. The first example demonstrates the 3D structures that can be created in GASS and the second one intends to reproduce an already existing model published in de Gregorio-Monsalvo et al. (2013).

### 6.1 Example 1 – 3D demonstration

The $^{13}$CO J = 2 → 1 emission of an object composed of a spherical source (assimilated to a protostellar envelope), a protoplanetary disc and a bipolar outflow is modelled in this example. Table 3 summarizes the physical properties set in this example, which does not correspond to any already observed case. All the structures are located in the centre $(0, 0, 0)$ of the model and the resulting complete velocity field given by these structures is taken into account. We have chosen $^{13}$CO as an example because its transitions are not too optically thick compared to $^{12}$CO. For the disc model, the abundance drops a lot when the temperature is below 27 K, to reproduce the freeze-out of $^{13}$CO on to dust grains. This will highlight clearly the emission regions of the disc for this example.

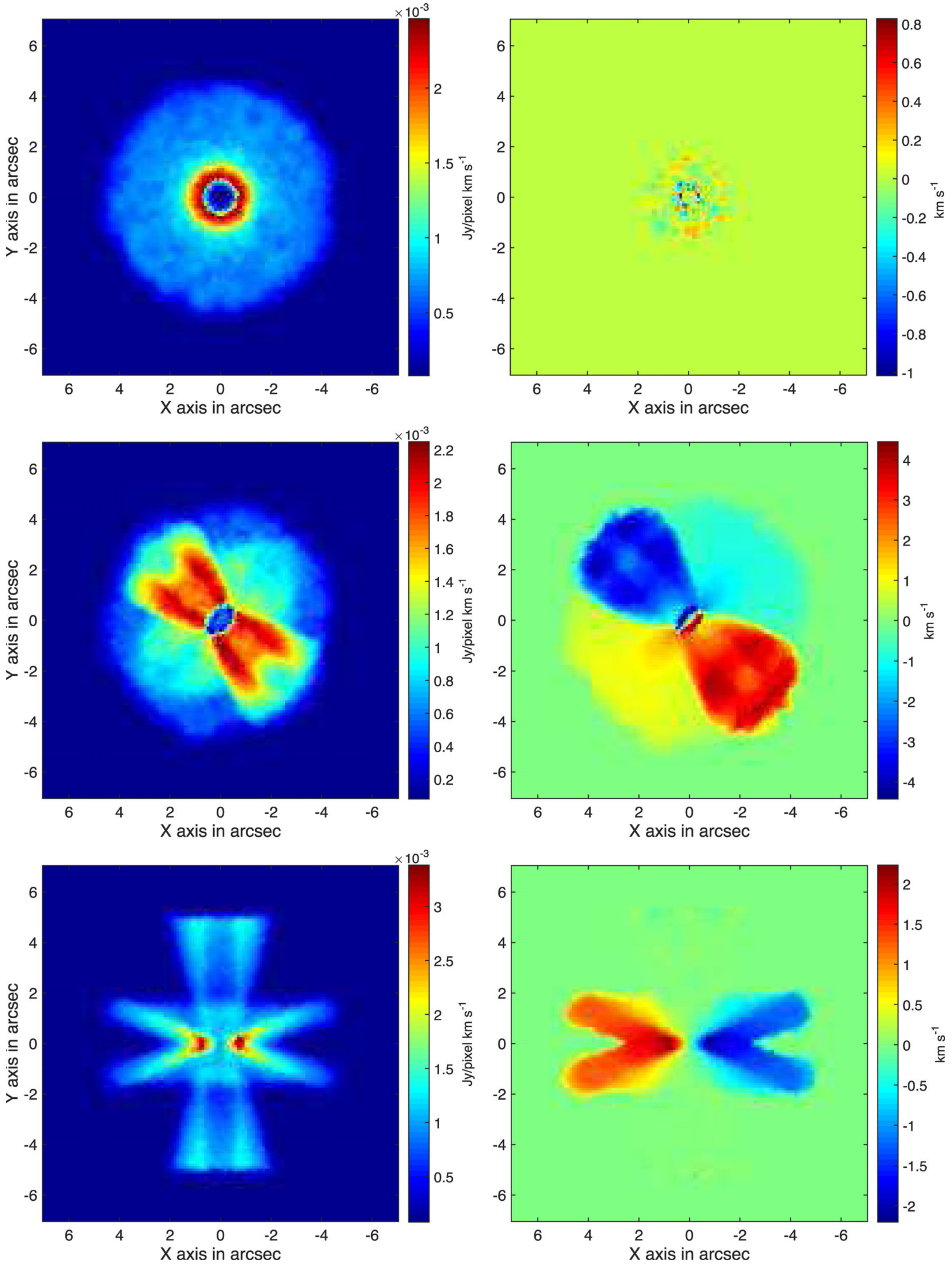

**Figure 13.** Moment 0 (left) and moment 1 (right) maps calculated over all channels of the data cubes for the $^{13}$CO J $= 2 \rightarrow 1$ transition. The model has been rotated by different angles: $\theta = 0°$, $\phi = 0°$ (top panels); $\theta = 45°$, $\phi = 45°$ (middle panels); $\theta = 90°$, $\phi = 90°$ (bottom panels).

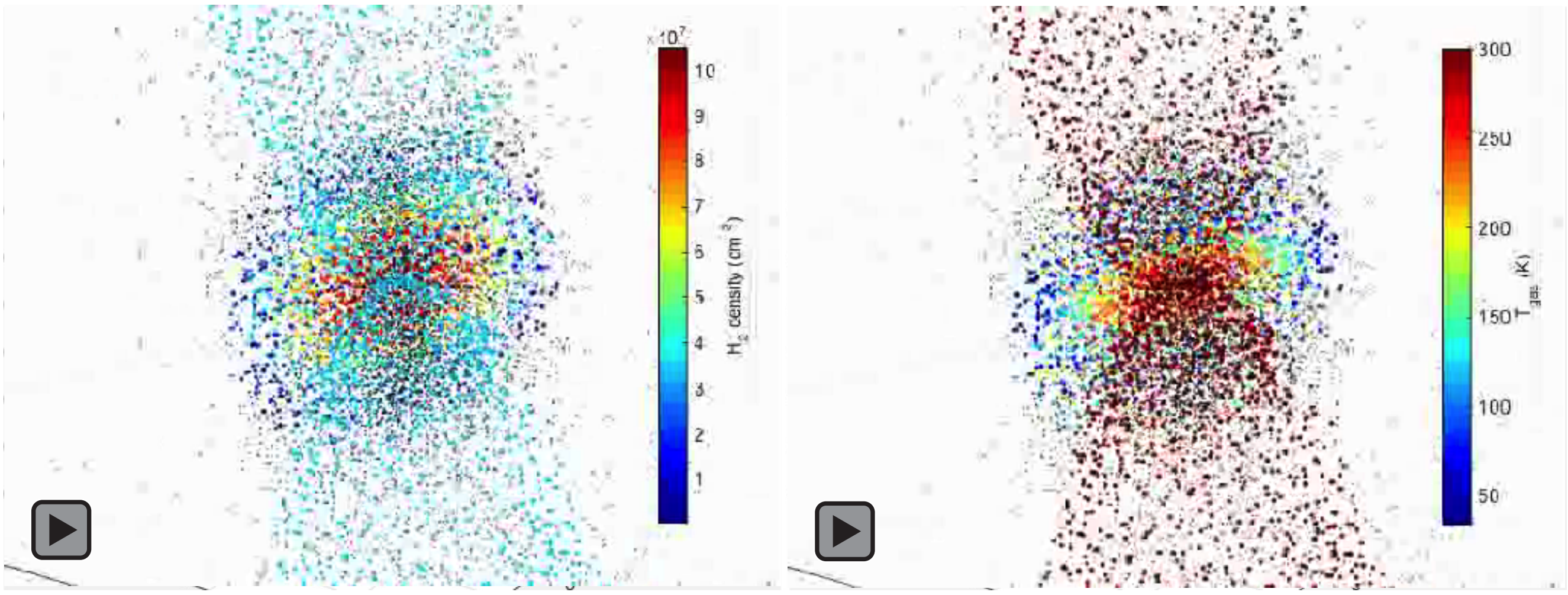

**Figure 14.** Animation of the 3D density (left) and gas temperature (right) structure of the first example. The outflow and disc structures are clearly recognisable. Click on the figure to activate the movie. Only works with Adobe Acrobat Reader, version ≥9 (not greater than 9.4.1 on Linux) or Foxit Reader.

The output data cube was set with 151 channels and a spectral resolution of 100 m s$^{-1}$ to cover a 15 km s$^{-1}$ bandwidth. The spatial resolution is 0.1 arcsec with 141 pixels to cover the radial size of 7 arcsec of the model. From Appendix B4, one can note that the pixel size is small enough to cover the physical variation of the example. The same physical case considering the dust continuum would have required a pixel size of 0.01 arcsec.

Thanks to GASS, we ran 10 times the same model and averaged them as described in Section 5.1 in order to completely blur out the artefacts due to the gridding process. From the fits file of the smoothed data cubes, Fig. 13 shows the resulting moment 0 and moment 1 maps over all channels for different sets of the ($\theta$, $\phi$) angles. In the moment 0 map, the outflow cavity emission is plainly identifiable with the $\theta = 0°$, $\phi = 0°$ (top-left panel, face-on) and $\theta = 90°$, $\phi = 90°$ (bottom-left panel, edge-on) models. The disc emission is also well identifiable at the centre of the image, especially in $\theta = 90°$, $\phi = 90°$ (bottom panels), where the disc is completely seen edge-on. The depletion of $^{13}$CO is well marked as expected. In the moment 1 map, the outflow is clearly identified in the middle-right panel thanks to its ejection velocity. The Keplerian rotation of the disc is also present, but fainter. In the bottom-right panel, this rotation is dominating, and only the disc is visible. The top-right panel does not show any sign of structures since the positive and negative velocity components cancel each other out along the line of sight.

The 3D structure of a model with $\theta = 45°$, $\phi = 45°$ is animated in Fig. 14 (click on the figure to activate the animation). The animation progressively zooms in and out of the 3D structure, displaying the $H_2$ density on the left-hand panel and the gas temperature on the right-hand panel. For a better 3D visualization, it is needed to slightly change some of the parameters of the disc and outflow structures. Therefore, the model presented in these animations is not exactly the same as the one described in Table 3. From these animations, one can note that the outflow cavity is clearly defined as well as the protoplanetary disc. All the points surrounding these two structures belong to the envelope model.

### 6.2 Example 2 – TW Hya

For this second example, we aim at reproducing the CO J = 2 → 1 ALMA Science Verification observations of the TW Hya protoplanetary disc. This disc has been well studied in the literature and several standard models are available (e.g. Thi et al. 2010; Andrews et al. 2012; Rosenfeld et al. 2012) so it is a good candidate to test GASS with a real astrophysical source. This example is based on both the ‘fiducial’ and ‘high-$q$’ models presented in Rosenfeld et al. (2012) (see their table 2 for the input parameters). We have selected the fiducial model because, as mentioned in section 3.2 of Rosenfeld et al. (2012), it is similar to the Model sA of Andrews et al. (2012), which is the only one providing a good match to the observed CO emission. The high-$q$ model was arbitrarily chosen among the alternative models of Rosenfeld et al. (2012) to check the effect of the variation of the parameters.

For each model, we have used GASS by giving an input file containing the density and temperature of the disc, as explained in Section 3.2. Once LIME has performed the radiative transfer calculations, we have used GASS to convolve the output data cube by the ALMA beam sizes. Information about the ALMA Band 6 observations such as the beam size has been taken from Rosenfeld et al. (2012).

Figs 15 and 16 present, respectively, the results of the fiducial and high-$q$ models compared to the ALMA observations. The high-$q$ model gives a better fit to the observations than the fiducial model, similarly to results presented in Rosenfeld et al. (2012).

We managed with GASS and LIME to reproduce the results obtained by two models of the study of Rosenfeld et al. (2012), showing that GASS is capable of performing state-of-the-art modelling of discs.

## 7 CONCLUSIONS

We have developed GASS, a code that allows us to easily define the physical structure of different astrophysical structures by creating, manipulating and mixing several different physical components such as spherical sources, discs and outflows (see Fig. C1). GASS can create input model files for LIME and the output data cubes generated by LIME can be analysed by several post-treatment options in GASS such as plotting spectra, moment maps or simulating observations. One must take care of LIME input parameters (number of points, pixel size) set for a given physical case since the resulting data cube may not be representative if they are not carefully chosen. We will keep on working on the development of GASS and, for instance, we expect to improve the GUI or to implement more LIME options in future releases of the code.

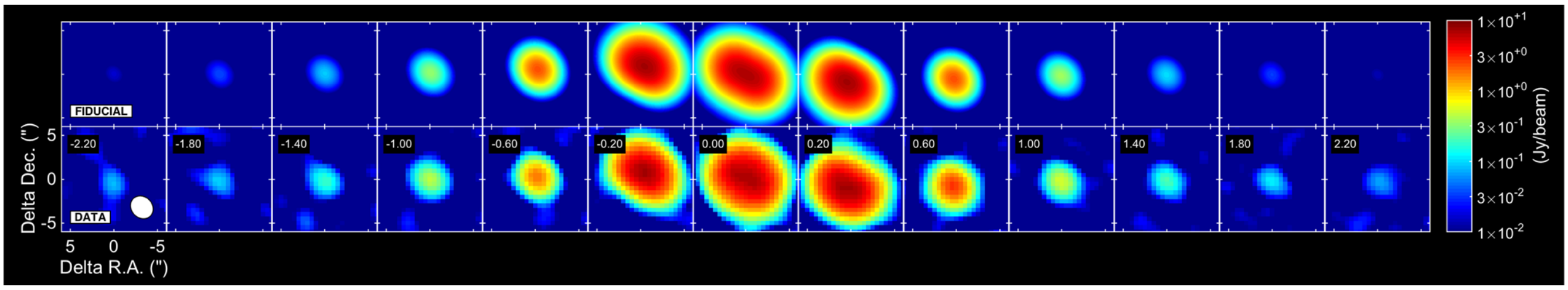

**Figure 15.** Channel maps of the data (bottom panels) compared to the fiducial model (top panels) of CO J = 2 → 1 emission, centred on the rest frequency of the transition. The channel spacing is 0.20 km s$^{-1}$. The ALMA beam is shown in the bottom right of the first panel.

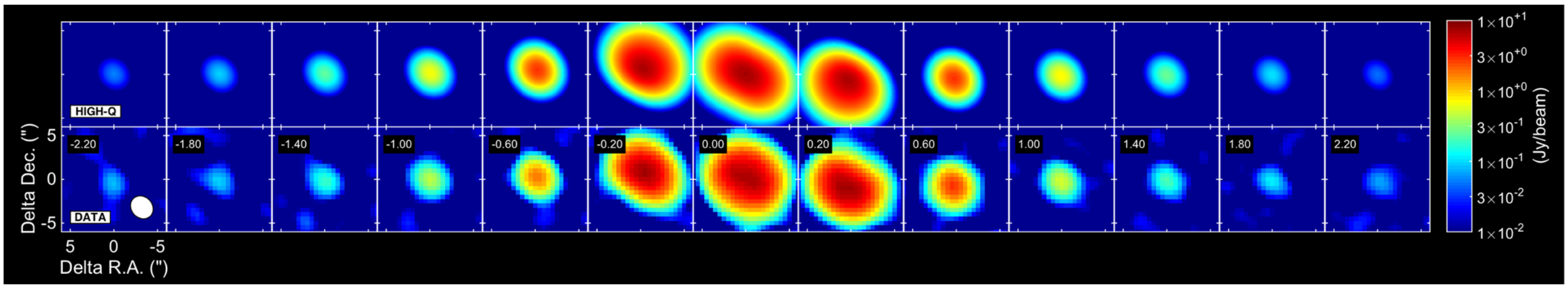

**Figure 16.** Same as Fig. 15 but for the high-*q* model.

## ACKNOWLEDGEMENTS

The authors would like to acknowledge Michiel Hogerheijde and Ian Stewart for their advices about the comparison between LIME and RATRAN. We also acknowledge Christian Brinch who gave precisions about the gridding process in LIME and Olivier Berné for very useful discussions about the disc modelling. We are thankful to the referee for the extremely useful and detailed comments on the paper. This paper makes use of the following ALMA data: ADS/JAO.ALMA#2011.0.00001.SV. ALMA is a partnership of ESO (representing its member states), NSF (USA) and NINS (Japan), together with NRC (Canada), NSC and ASIAA (Taiwan) and KASI (Republic of Korea), in cooperation with the Republic of Chile. The Joint ALMA Observatory is operated by ESO, AUI/NRAO and NAOJ.

## APPENDIX A: DELAUNAY TRIANGULATION AND VORONOÏ DIAGRAMS

Both GASS and LIME are using the Voronoï diagram to build the computational grid. Indeed, from a random grid of points it is possible to build both a Voronoï diagram or a Delaunay triangulation (Delaunay 1934) and in this appendix we will explain how both are generated. By construction, these two objects are linked to each other. The Delaunay triangulation is generated by connecting three neighbouring points of the grid. These points define the Delaunay circle and no other point of the grid lies in the circle. From the Delaunay grid, we can construct the Voronoï cells (Voronoï 1908). Fig. A1 shows a sketch of how the Voronoï diagram is built from the Delaunay triangulation of the black dots. The three bisectors (in blue) of each Delaunay triangle (in red) define the centre of the circumcircle (in brown) of each triangle. This centre defines a vertex (green points) of the Voronoï diagram. Thus, each bisector of a Delaunay triangle is an edge of a Voronoï cell.

In GASS, only the Voronoï diagram is used to build and smooth the grid. However, LIME is building both of them from the distribution of points given by GASS. This could seem redundant but the current version of LIME does not allow us to give directly in input the Voronoï cells (nor the Delaunay grid).

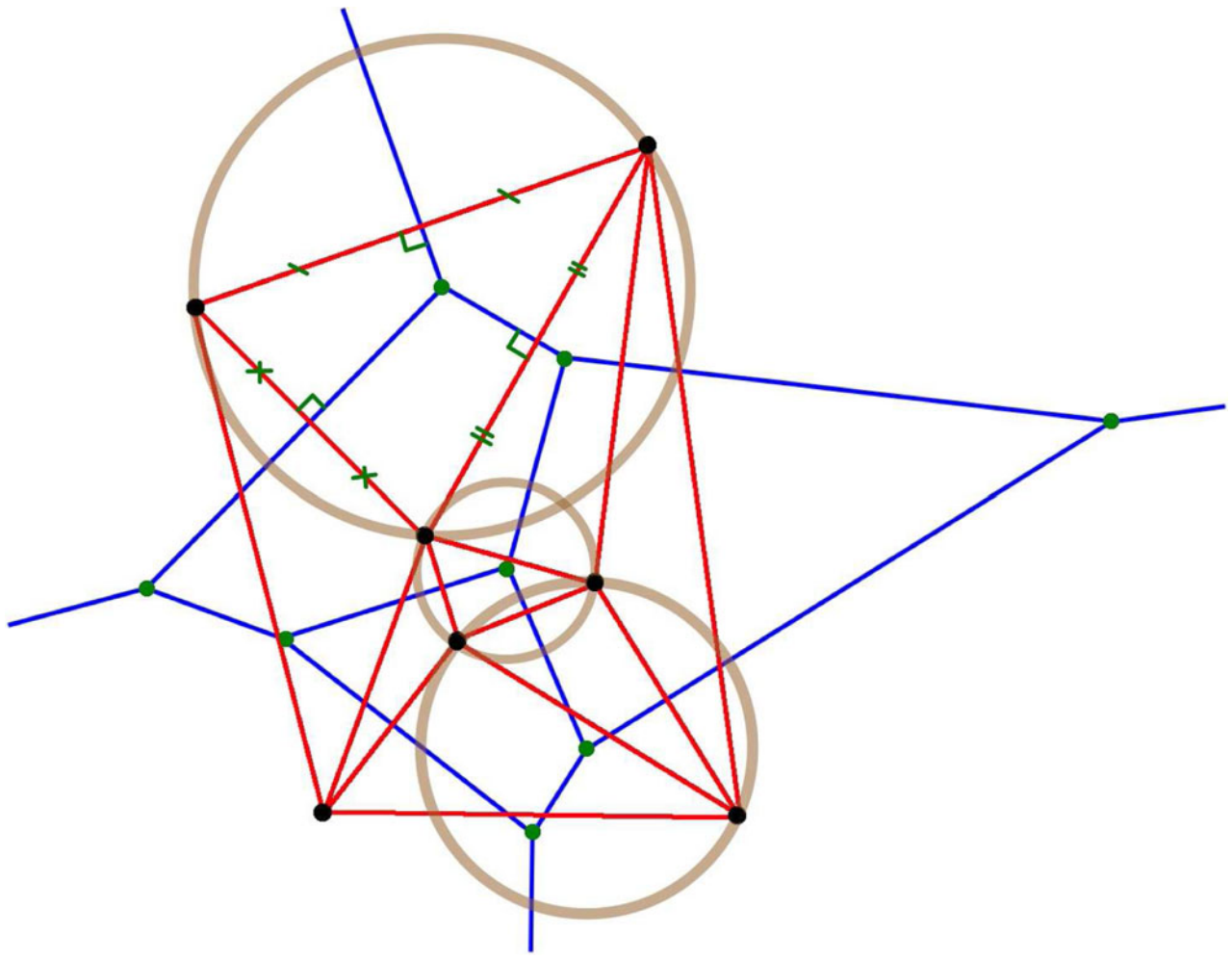

**Figure A1.** Example of the Voronoï cells building. The black dots are the grid points and the red lines show the Delaunay triangulation from which the Voronoï cells are derived in blue. Three examples of the Delaunay triangle circumcircle are plotted in brown. One example of the three bisector construction in one Delaunay triangle is also displayed.

The Delaunay triangulation is only used in GASS to calculate the total gas mass of a given disc (see Section 3.2). In this case, the volume of each tetrahedron generated by the Delaunay triangulation is needed. This volume is calculated from the coordinates of the four vertices of a given tetrahedron. If we consider a tetrahedron with vertices $\boldsymbol{a}$, $\boldsymbol{b}$, $\boldsymbol{c}$, $\boldsymbol{d}$, its volume $v$ is given by

$$V = \frac{|(\boldsymbol{a}-\boldsymbol{d})\cdot((\boldsymbol{b}-\boldsymbol{d})\times(\boldsymbol{c}-\boldsymbol{d}))|}{6}. \quad \text{(A1)}$$

This volume is then used in combination to the density to obtain the mass of each tetrahedron.

## APPENDIX B: IMPACT OF LIME PARAMETERS

In this section, we aim at showing the impact of different input parameters of LIME on the resulting data cubes and confront them to the ones given by RATRAN. The goal is to show that badly chosen input parameters of LIME can lead to wrong results, independently of the radiative transfer calculation. Therefore, the users are strongly advised to take great care in choosing their parameters in order to avoid these potential problems.

To perform these tests, we used the RATRAN 1D version of 2013 March and the LIME version 1.5. LIME has already been benchmarked against RATRAN (Brinch & Hogerheijde 2010) using the problem and solution defined by van Zadelhoff et al. (2002), so we can compare one to the other as long as we stay in the 1D regime. Indeed, since we used the 1D version of RATRAN, we can only consider structures with a spherical symmetry centred at the origin of the grid.

We have run several models to test the impact of the input parameters of LIME on the resulting image. The comparison with RATRAN is based on:

(i) the population density of the energy levels as a function of the radius (see Section B1);

(ii) the shape and intensity of the line profile, as well as the value of the predicted continuum level, from different beam positions and sizes in the map (see Section B3).

**Table B1.** 'Benchmarking' model properties.

| | |
|---|---|
| Common parameters | |
| Number of channels | 71 |
| Channel resolution | 100 m s$^{-1}$ |
| Image size | 171 × 171 |
| Pixel size | 0.2 arcsec |
| Outer radius | 6000 au (50 arcsec at 120 pc) |
| Gas-to-dust ratio | 100 |
| Code-specific parameters | |
| RATRAN shell numbers | 191 |
| LIME number of points | 101 992 |

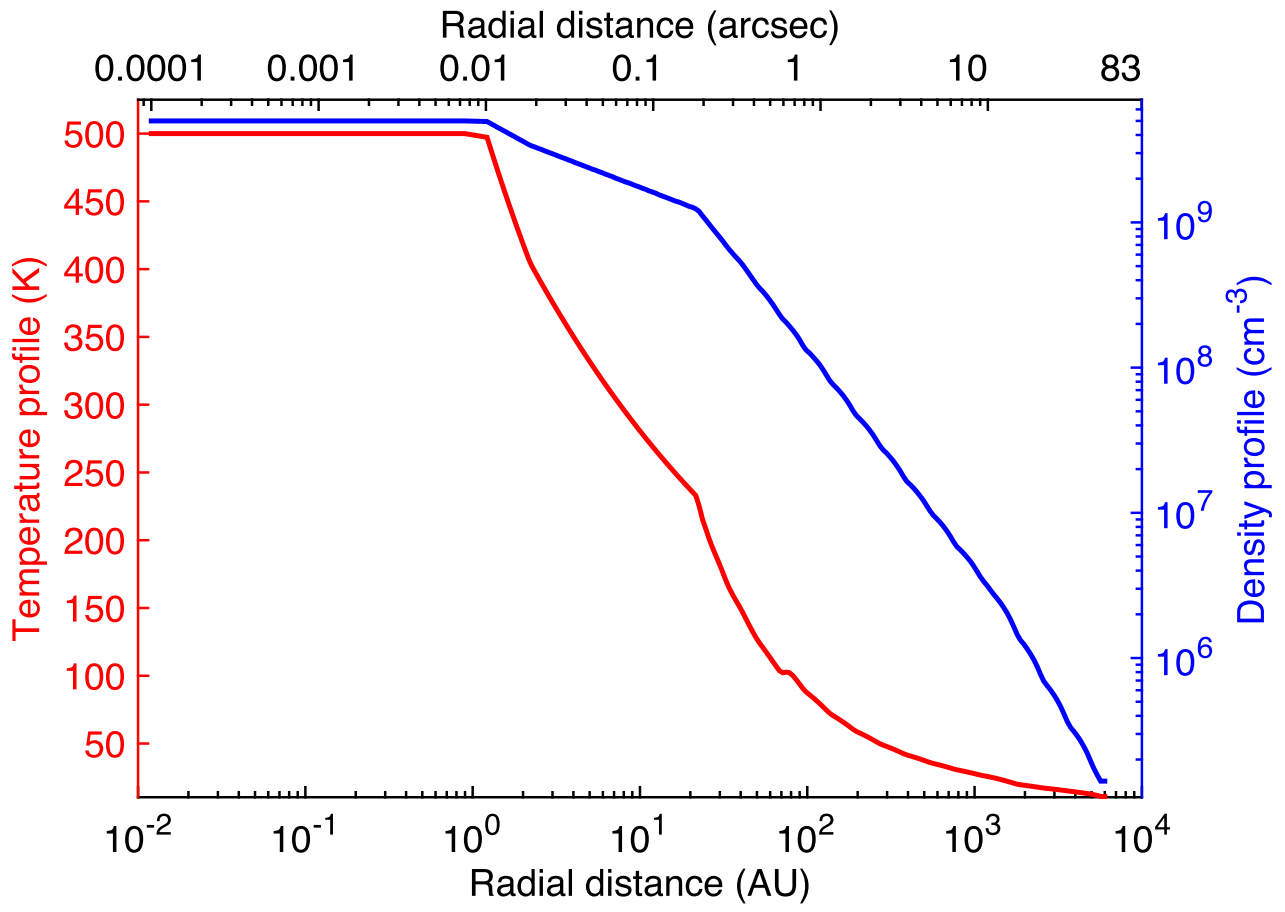

**Figure B1.** Gas and dust temperatures (in red) and $H_2$ density (in blue) as a function of the radius.

### B1 Physical model

The model is based on the physical structure of the low-mass protostar IRAS16293−2422 as derived by Crimier et al. (2010): a collapsing spherical source around a central object located at 120 pc. The emission of $HCO^+$ from $J = 1 \rightarrow 0$ to $J = 13 \rightarrow 12$ is computed. The input parameters for LIME and RATRAN used in this study are listed in Table B1. Variable gas and dust temperatures as well as $H_2$ density profiles are used as a function of the radius (see Fig. B1) and we set a constant abundance and $b$-Doppler value all over the model of $5 \times 10^{-12}$ and 200 m s$^{-1}$, respectively.

By construction, in the two codes, the velocity field plays an important role in the resulting data cubes. In a spherical model, we have verified that LIME and RATRAN give exactly the same result if the velocity field in RATRAN is given as a function rather than being written on the grid (Hogerheijde & van der Tak 2000). For the purpose of this study, the velocity field does not play an important role, and as suggested by the authors of both LIME and RATRAN (Hogerheijde et al., private communication) we decided to set the velocity field to zero for this study.

### B2 Population density of energy levels

The population density of the different energy levels of the molecule given by LIME and RATRAN is shown in Fig. B2. There is a good agreement between RATRAN and LIME for the calculation of the population density of the first five levels of $HCO^+$. One can note that the LIME curves become a bit ratty at large radii, this is due to the cell density becoming smaller in the outer part of the model. This effect does not affect the resulting images since the mean values stay very close to the RATRAN ones.

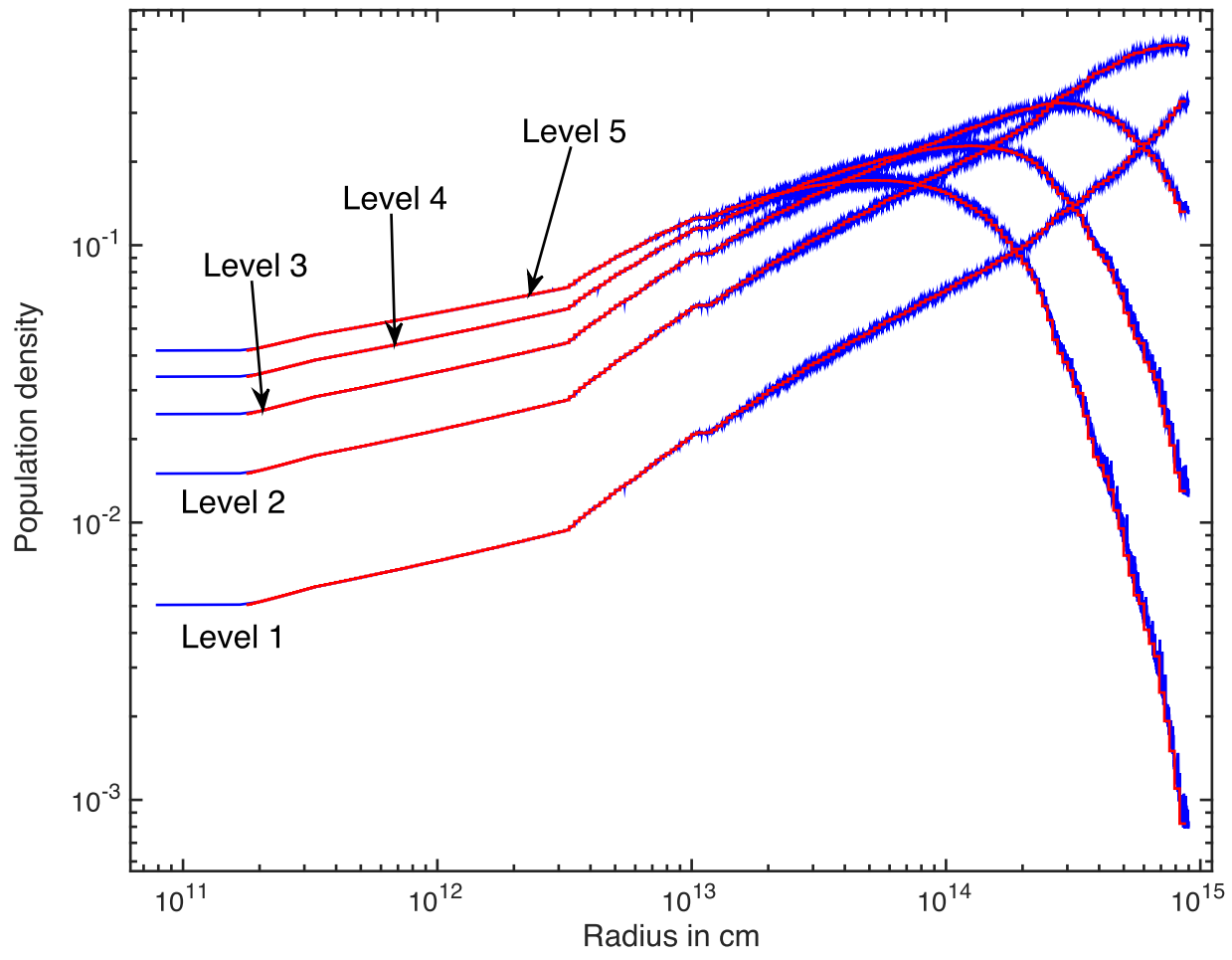

**Figure B2.** Population density of the first five levels of $HCO^+$ as a function of the radius. The blue curves correspond to LIME and the red ones to RATRAN.

## B3 Line profiles and continuum levels

The goal here is to investigate the difference between the line profiles and continuum levels calculated by RATRAN and LIME in order to find some 'diagnostics' allowing us to illustrate the effects of some of the LIME input parameters. To do so, we choose to look at three transitions among all the calculated ones, namely the J = 1 → 0, J = 6 → 5 and J = 13 → 12 transitions. These transitions span a wide range of upper energy levels, $E_{up}$ (∼4 K, ∼90 K and ∼400 K, respectively), hence probe different physical conditions, and so they represent a good sample to trace the differences between the two codes.

Figs B3a, B3b and B3c display the J = 1 → 0, J = 6 → 5 and J = 13 → 12 transitions, respectively. In each of the three figures, the left-hand panel shows the resulting image of the difference Diff in per cent between the run of RATRAN and the run of LIME, calculated by

$$\text{Diff[per cent]} = 100 \times \frac{\text{RATRAN} - \text{LIME}}{\text{RATRAN}}. \quad \text{(B1)}$$

To perform the comparison, we define circular regions centred at three positions, [0, 0] arcsec, [0, −11.6] arcsec and [−11.6, 11.4] arcsec and with radii $R = 1.0, 3.3, 5.6, 7.9, 10.2$ arcsec (at the central position) and $R = 1.0, 3.3, 5.6$ arcsec (at the outer positions). For each circular region (loosely called 'beam' hereafter), we compute the average of Diff: except for the J = 1 → 0 transition with beams centred on [0,0] arcsec, we find that the absolute value of this average, in any other combination of beam size and position, is less than ∼6 per cent. We also compute the average spectrum in each beam, i.e. the mean value of all the pixels contained in that beam: the average spectra for three of the considered beam sizes and positions ($R = 1$ arcsec and $R = 5.6$ arcsec at [0, 0] arcsec and $R = 5.6$ arcsec at [−11.6, 11.4] arcsec – see the black circles in the left-hand panels of Fig. B3) are displayed in the right-hand panels of Fig. B3.

We chose these three beams in order to illustrate the following trends for the J = 6 → 5 and J = 13 → 12 transitions (Figs B3b and c).

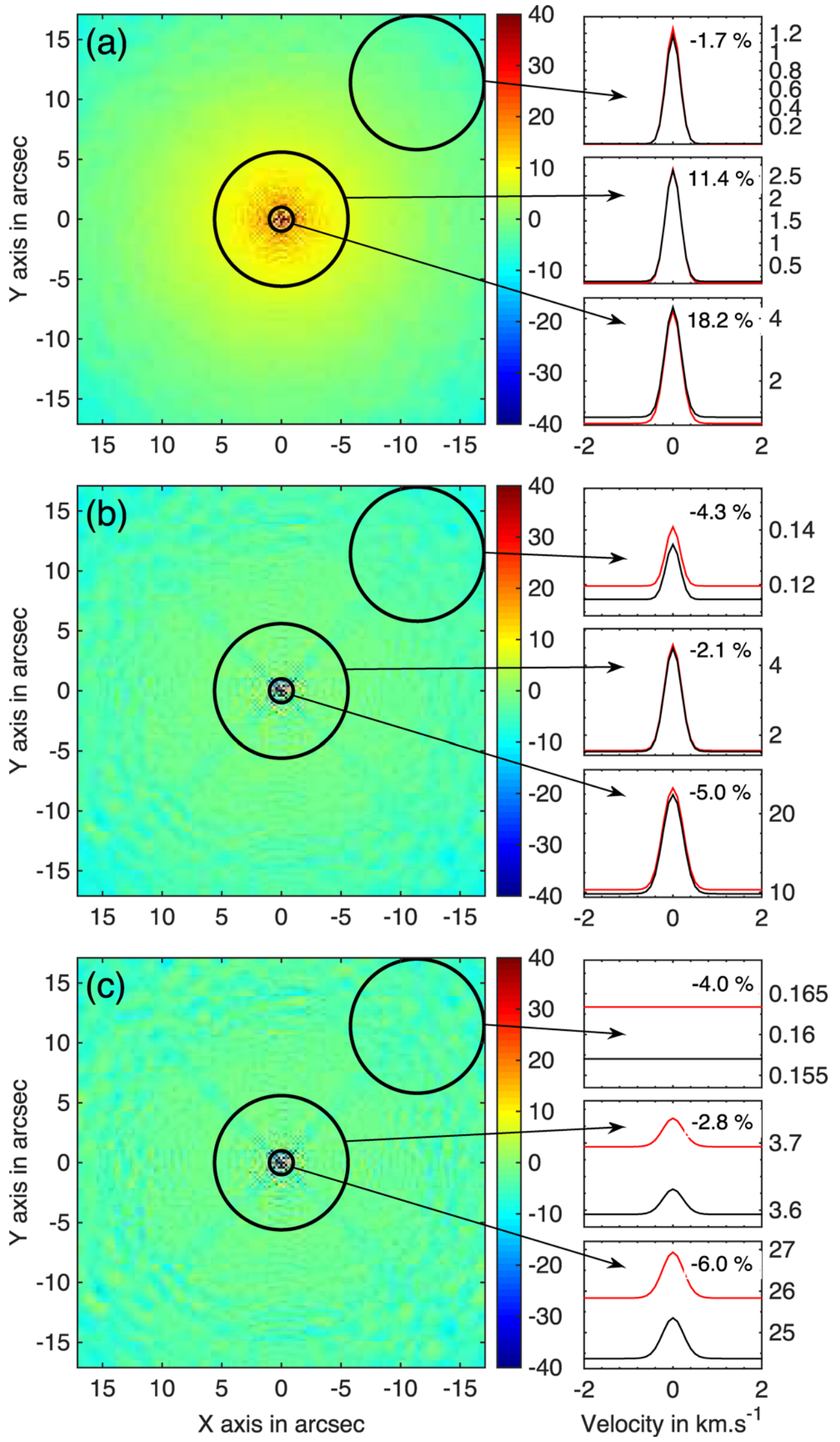

**Figure B3.** Left-hand panels: comparison in per cent (see equation B1) between the output cubes of RATRAN and LIME for the $HCO^+$ J = 1 → 0 (panel a), J = 6 → 5 (panel b) and $J = 13 \rightarrow 12$ (panel c) transitions. The black circles show the positions and sizes of the beams used to compute the spectra: $R = 1$ arcsec and $R = 5.6$ arcsec at [0, 0]arcsec and $R = 5.6$ arcsec at [−11.6, 11.4]arcsec. Right-hand panels: RATRAN (black) and LIME (red) spectra (in K) averaged over the corresponding regions of the left-hand panels. The averaged difference in per cent is indicated in the upper right corner of each panel.

(i) Decreasing difference (in absolute value) with increasing radius as shown by the spectra corresponding to $R = 1$ arcsec and $R = 5.6$ arcsec at [0, 0] arcsec in Fig. B3b and c. First, by looking at the left-hand panels of Fig. B3, one can note that the innermost part of the images ($R \lesssim 1.5$ arcsec) shows a pixel-to-pixel difference larger than in the outer part of the images. In this central region, the pixel-to-pixel difference (in absolute value) can reach a value of up to 40 per cent in contrast to the average value of $\lesssim$6 per cent (excluding the J = 1 → 0 transition) obtained when averaging in a beam, even with a radius as small as 1 arcsec. We show in Appendix B4 that this large pixel-to-pixel difference is due to the different continuum levels calculated by RATRAN and LIME, which are noticeable in the spectra plotted in the right-hand panels of Fig. B3 (one must take care of the different $y$-axis scales between the spectra). This large pixel-to-pixel difference is washed out when averaging over the central 2 arcsec, and the absolute difference decreases further when averaging over a larger number of pixels.

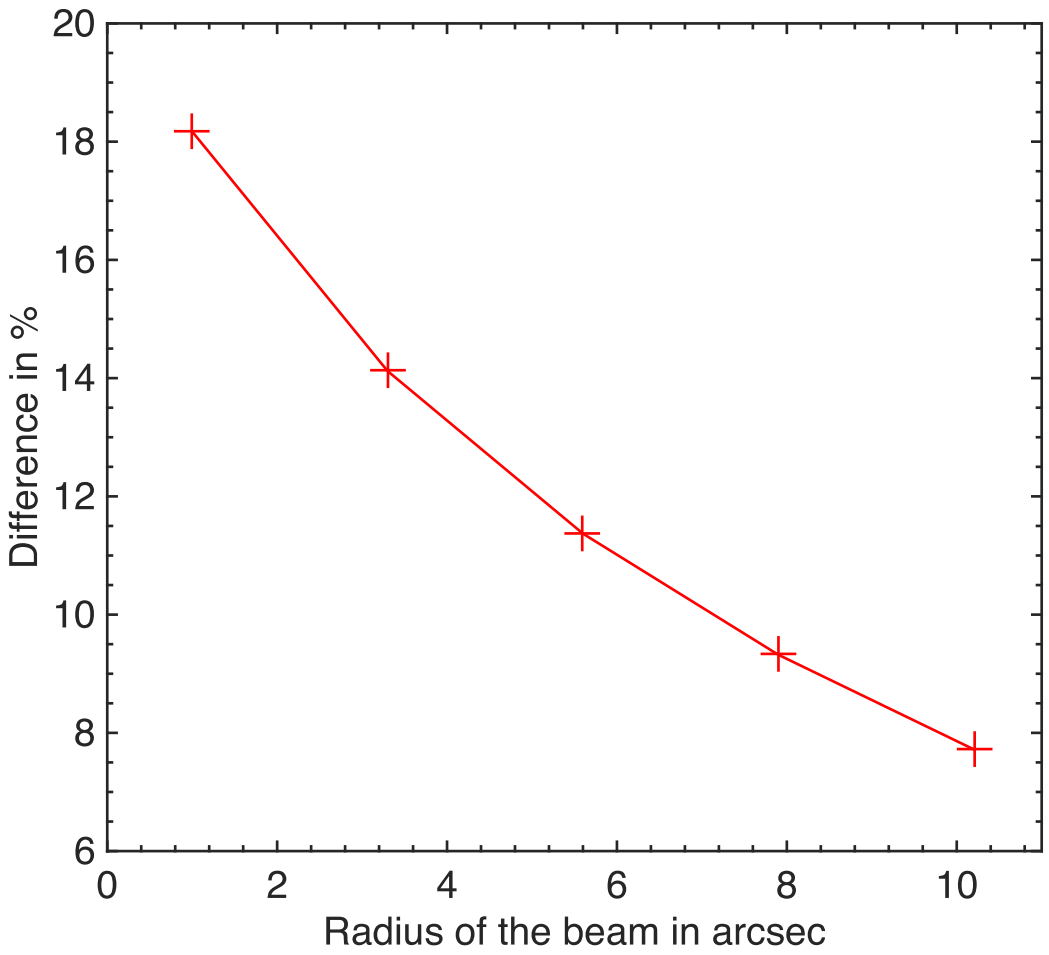

**Figure B4.** Difference in per cent (see equation B1) between the LIME and RATRAN maps as a function of the beam size in arcsec, centred at [0, 0] arcsec, for the $J = 1 \rightarrow 0$ transition.

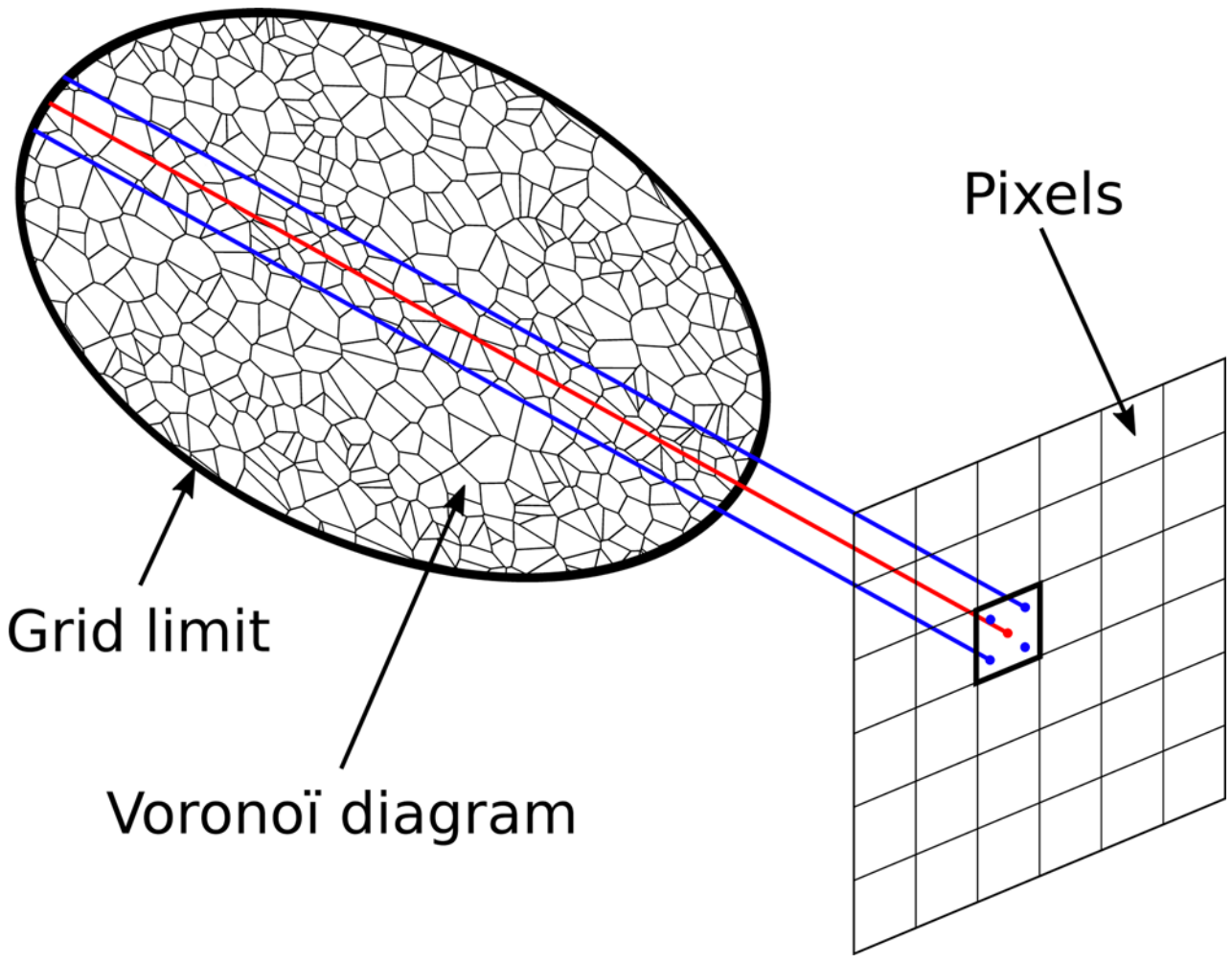

**Figure B5.** Sketch explaining the ray-tracing of LIME. The pixel intensity is only determined with cells crossing the different lines of sight. The red line is the line of sight located at the centre of a pixel, for *anti-alias* = 1. Blue ones show the distribution of lines of sight for *anti-alias* = 4.

(ii) Increasing difference (in absolute value) with increasing distance from the centre as shown by the spectra corresponding to $R = 5.6$ arcsec at [0, 0] arcsec and at [−11.6, 11.4] arcsec in Fig. B3b and c: we show in Section B5 that this trend can be reduced by adopting a different point distribution.

Fig. B4 shows the average of Diff as a function of beam radius for the central position for the $J = 1 \rightarrow 0$ transition. As mentioned above, for all the other combinations of transition/beam position/beam size, the absolute averaged difference is less than ∼6 per cent, hence, for the sake of clarity, we do not plot them in Fig. B4. As for the other transitions, this difference could be due to the different continuum levels. However, after investigation (see section B4), it turns out that this effect plays a little role in the large averaged difference observed in this case. Indeed, the $J = 1 \rightarrow 0$ transition of $HCO^+$ transition has a low $E_{up}$ (∼4 K), leading to a particularly extended emission compared to the other transitions. Since this line is highly optically thick ($\tau \sim$ few hundreds versus ∼20 and ∼1 for the $J = 6 \rightarrow 5$ and $J = 13 \rightarrow 12$ transitions respectively), opacity effects resulting from the *ray-tracing* process are summing up in a much larger region, leading to a higher difference between the two codes.

## B4 Impact of the pixel size

In this section, we show the impact of the pixel size in the continuum calculation. Indeed, we initially use a pixel size of 0.2 arcsec. However, the physical properties of the described model are varying a lot in a region that corresponds to this pixel size: $n(H_2)$ decreases by about a factor of 10, and, most importantly, the temperature drops by ∼300 K. Since the dust emission is strongly dependent on the temperature profile, the continuum will strongly vary from one pixel to the next.

Now, to determine the pixel intensity, LIME ray-traces photons in straight lines, considering a certain number of lines of sight per pixel (typically 1–4), controlled by the value of the optional parameter *anti-alias*. LIME uniformly distributes these lines of sight over the pixel surface, as can be seen in Fig. B5, and the intensity in the pixel is the average of the different lines of sight. RATRAN calculates the intensity in a slightly different way but the pixel-to-pixel difference can also be large. So when the difference goes into opposite directions between the two codes, it leads to an even higher value of Diff.

If the *anti-alias* option of LIME is set to 1 and if the pixel size is greater than the mean dimension of cells located behind, then only the physical properties of cells located at the pixel centre will be taken into account, distorting the final results (and leading to a large difference compared to RATRAN). To avoid this issue, one must decrease the pixel size or increase the value of the *anti-alias* option. In any case, it will increase (almost in a similar way) the executing time of LIME.

We therefore performed the models with a pixel size of 0.02 arcsec with a number of pixels increased from 171 to 1701 to maintain the same map size (all other parameters of Table B1 stay the same) and illustrated the effect in Fig. B6. The middle panel of this figure displays the $J = 13 \rightarrow 12$ transition modelled with a pixel size of 0.02 arcsec. To better see the effect of this reduced pixel size, only an inner map of 10 arcsec × 10 arcsec, compared to the initial ∼34 arcsec × 34 arcsec map shown in Fig. B3c, is displayed. To facilitate the comparison, we also display in the left-hand panel of Fig. B6 a zoom of Fig. B3c showing the same central region as the middle panel: the pixel-to-pixel difference in the middle panel is much smaller than in the left-hand panel. The improvement is also noticed in the spectra averaged over an $R = 1$ arcsec circle and displayed in the right-hand panel of Fig. B6 (compared with the right, bottom panel of Fig. B3c): the absolute difference is now only 1 per cent, versus 6 per cent previously. A similar improvement is found for all the other transitions, except for the $J = 1 \rightarrow 0$ transition: this indicates that the difference in the continuum levels plays a little role for this transition and that the large difference we observe is dominated by some other effect(s) (see section B3).

## B5 Impact of the point distribution

The importance of the point distribution on the difference has also been tested. The exact same model has been used considering a linear distribution of points as a function of the radius instead of a distribution following the density as a function of the radius (see Section 2). A linear distribution produces more points in the external part of the model compared to the density distribution, considering the density profile shown in Fig. B1. With this new distribution, the difference is reduced by a factor of up to 2, depending on the beam radius and position. The gain is greater on the outer part of the image and for the smaller beam radii. This trend shows that the point distribution, as well as the number of points, are both important parameters to consider while working with LIME. However, this new distribution does not change much the final result given by LIME because the difference was already small ($\lesssim$6 per cent) for the previous distribution; but it explains why the difference plotted in Fig. B4 does not decrease as much as one might have expected.

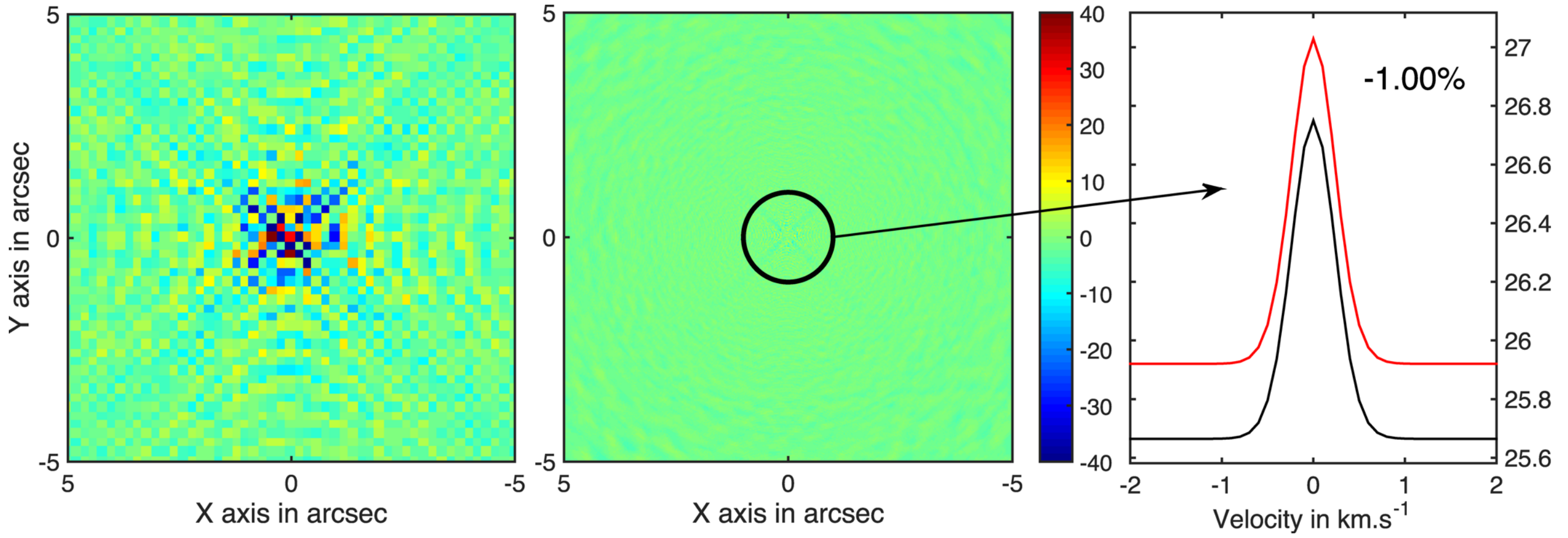

**Figure B6.** Comparison in per cent (see equation B1), for the central 10 arcsec between the output cubes of RATRAN and LIME for the $HCO^+$ J = 13 → 12 transitions modelled with a pixel size of 0.2 arcsec (left-hand panel) and 0.02 arcsec (middle panel). The black circle shows the $R = 1$ arcsec beam used to compute the spectra displayed in the right-hand panel. The averaged difference in per cent is indicated in the upper right corner of this panel.

### B6 Summary

In conclusion, depending on the science goals, great care must be taken when choosing the input parameters for LIME, in particular:

(i) the pixel size should be small enough to correctly probe the steepest gradient of the physical model over the scale that the user is interested in. Alternatively, if a larger pixel size is used, the user should increase the *anti-alias* parameter;

(ii) if the user is interested in the emission at the edges of the map, s/he should consider a point distribution yielding more points in the external part of the model.

## APPENDIX C: GASS ANIMATION

In this appendix, a 3D model generated by GASS is displayed, combining multiple structures in a complex physical model. Fig. C1 displays one outflow, two discs and two spherical sources.

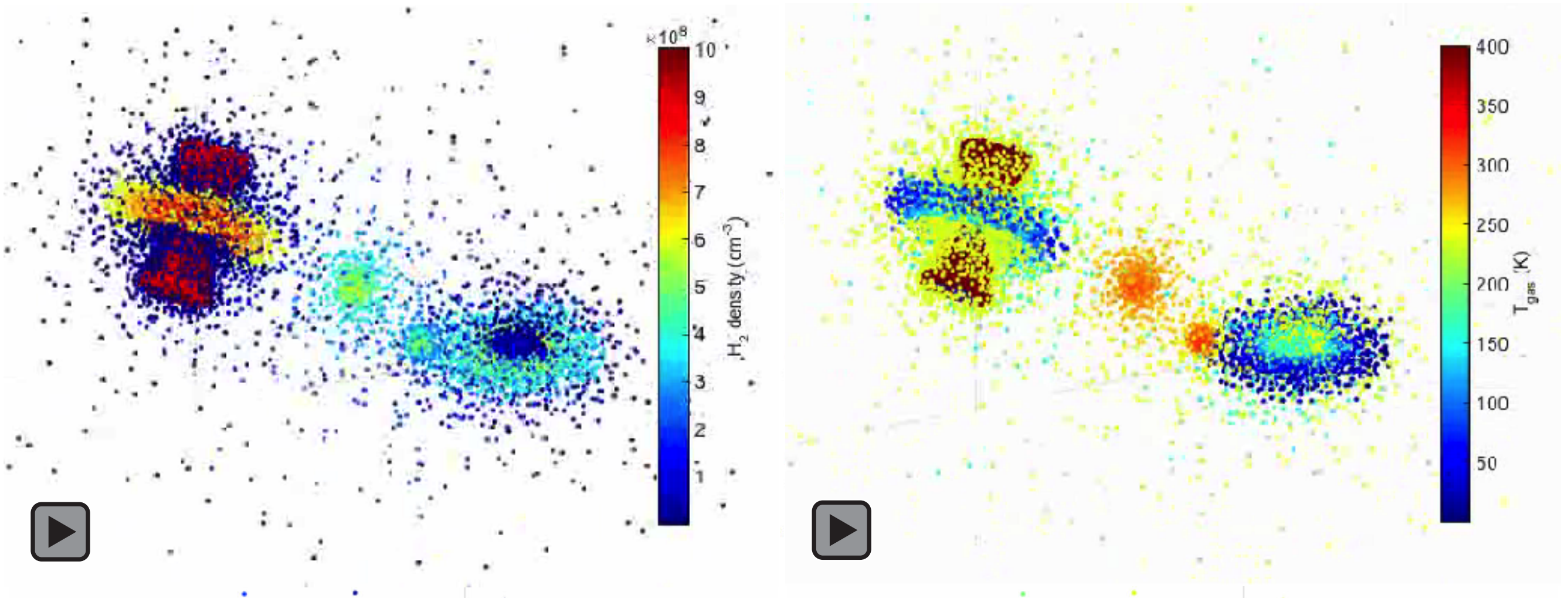

**Figure C1.** Animation of the 3D density (left) and gas temperature (right) structure of a complex model, containing two discs, two spherical cores and one outflow. Click on the figure to activate the movie. Only works with Adobe Acrobat Reader, version ≥9 (not greater than 9.4.1 on Linux) or Foxit Reader.